\documentclass[trackchanges]{aastex7}
\usepackage{amsmath}

\newcommand\bb[1]{\mbox{\boldmath{$#1$}}}
\newcommand\grad{\bb{\nabla}}
\newcommand\bcdot{\,\bb{\cdot}\,}
\newcommand\bdbldot{\,\bb{:}\,}
\newcommand\btimes{\,\bb{\times}\,}

\usepackage{hyperref}

\begin{document}

\title{Generation of proton beams at switchback boundary-like rotational discontinuities in the solar wind}

\author[orcid=0000-0001-7655-5000,sname='Lin']{Rong Lin}
\affiliation{School of Earth and Space Sciences, Peking University, Beijing, China}
\affiliation{Centre for mathematical Plasma Astrophysics, Department of Mathematics, KU Leuven, Leuven, Belgium}
\email{fakeemail1@google.com}  

\author[0000-0002-7526-8154]{Fabio Bacchini} 
\affiliation{Centre for mathematical Plasma Astrophysics, Department of Mathematics, KU Leuven, Leuven, Belgium}
\affiliation{Royal Belgian Institute for Space Aeronomy, Solar-Terrestrial Centre of Excellence, Brussels, Belgium}
\email{fakeemail2@google.com}

\author[0000-0001-8179-417X]{Jiansen He}
\affiliation{School of Earth and Space Sciences, Peking University, Beijing, China}
\email[show]{jshept@pku.edu.cn}

\author[0000-0001-5079-7941]{Luca Pezzini}
\affiliation{Centre for mathematical Plasma Astrophysics, Department of Mathematics, KU Leuven, Leuven, Belgium}
\affiliation{Solar-Terrestrial Centre of Excellence--SIDC, Royal Observatory of Belgium, Brussels, Belgium}
\email{fakeemail3@google.com}

\author[0000-0001-7537-5999]{Jingyu Peng}
\affiliation{School of Earth and Space Sciences, Peking University, Beijing, China}
\email{fakeemail4@google.com}

\correspondingauthor{Jiansen He}


\begin{abstract}

Alfvénic rotational discontinuities (RDs) are abundant in the inner heliosphere and can be used to model the boundary of switchbacks, i.e. Alfvénic magnetic kinks. To investigate the effects of RDs on proton kinetics, we model a pair of switchback-boundary-like RDs with a hybrid Particle-In-Cell (PIC) approach in a 2D system. We find that, at one of the boundary RDs, a significant population of protons remains trapped over long times, creating a secondary beam-like component with temperature anisotropy $T_\perp/T_\|\gtrsim4$ in the proton velocity distribution function that excites ion cyclotron waves within the downstream portion of the transition layer. Further analysis suggests that the static electric field in the vicinity of the RD is the key factor in trapping the protons. This work indicates that switchback boundaries could represent a viable environment for the creation of proton beams in the heliosphere; it also highlights the need to investigate RD sub-structures, especially the embedded current systems of interplanetary RDs. Finally, this paper underscores the importance of high-resolution observations of the solar wind velocity distributions around RDs.

\end{abstract}




\section{Introduction}
In solar wind physics, a proton beam is a secondary population of the proton velocity distribution function (VDF) \citep{Feldman1973, Marsch1982}. It is often characterized with an anti-sunward drift velocity relative to the primary population (the core). The drift velocity is generally aligned with the background magnetic field, and while its magnitude may occasionally exceed the local Alfvén speed, drift instabilities restrict this excess to remain small. 
Although the number density ratio of the beam to the core, $n_b/n_c$, is typically less than 0.5 at most times \citep{Marsch1982, Durovcova2021, Verniero2020}, the beams can still carry significant free energy and can potentially excite instabilities, thus modulating the local state of the solar wind \citep{gary_electromagnetic_1991,Hellinger2011,Verscharen2013}.
The generating mechanisms of proton beams have been proposed as Coulomb collision \citep{Livi1987}, wave--particle interaction with a second branch of right-hand polarized (RHP) and left-hand polarized (LHP) waves in the presence of alpha particles \citep{Tu2002}, parametrically unstable Alfvén-cyclotron waves \citep{Araneda2008}, trapping effect of obliquely propagating Alfvén modes \citep{Osmane2010, Durovcova2021}, and magnetic reconnection \citep{Lavraud2021}.

Rotational discontinuities (RDs) are one of the inherent types of magnetohydrodynamic (MHD) discontinuities formulated by the Rankine--Hugoniot relations \citep{landau1984electrodynamics}. They are characterized by a rotation of the magnetic field and plasma velocity, which are Alfvénically correlated, while plasma density and pressure stay nearly constant across the transition layer. RDs are commonly interpreted as the steepened forms of Alfvén waves \citep{Tsurutani1995,Vasquez2001,Yang2015} or as products of magnetic reconnection \citep{Lin1994,Lin2009,Liu2011,Innocenti2015}. The occurrence rate of RDs is positively correlated with the solar wind speed \citep{Burlaga1977,Neugebauer1984}. Dumbbell-shaped proton velocity distribution functions (VDFs) and $\alpha$ particle VDFs are observed in the reconnection exhaust region bounded by a pair of back-to-back RDs, which guide the particles crossing the RDs' interfaces at about Alfv\'en speed to form the dumbbell-shaped VDFs and cause the parallel heating \citep{he2018plasma, duan2023kinetic}. Moreover, enhanced Alfv\'enic turbulence full of bi-directional propagating Alfv\'en waves and the occurrence of oblique compressional waves are also observed in the reconnection exhaust region, as bounded by the back-to-back RDs \citep{he2018plasma, zhuo2024oblique}. Proton RDs are observed throughout interplanetary space \citep{Neugebauer1990, Neugebauer2006}. Numerical simulations of individual RDs with finite transition width of several ion gyroradii have focused on dominant mechanisms at the transition scale, addressing issues such as their stability, the minimum shear required for their existence, and their interaction and coexistence with background waves and fluctuations \citep{Richter1989, KraussVarban1993, KraussVarban1995, Karimabadi1995}. 

Parker Solar Probe (PSP) \citep{Fox2015} and Solar Orbiter (SolO) \citep{Mueller2020}, launched in 2018 and 2020, respectively, mark a new era of solar exploration and offer unique opportunities to revisit classic topics in space plasma physics such as proton beams and RDs. Equipped with the Solar Wind Electron Alpha and Proton (SWEAP) \citep{Kasper2015,Livi2022} and the Solar Wind Analyzer (SWA) \citep{Owen2020} suite of instruments, these missions provide observations of solar wind ion (primarily proton) VDFs at cadences down to $\sim 1$ second and 0.25 seconds, respectively, enabling detailed investigations of proton beams and related plasma instabilities \citep{Klein2021, Louarn2021}. Recent observations have further revealed novel proton VDFs, including three-component distributions where a third population, the ``hammerhead'', is strongly heated perpendicularly to the background magnetic field \citep{Verniero2022,Pezzini2024}, as well as non-field-aligned proton beams modulated by the fast magnetosonic/whistler waves excited by the proton VDF itself \citep{Zhu2023}.

Another important finding in the inner heliosphere is the prevalence of Alfvénic magnetic kinks, also called ``switchbacks'' \citep{Bale2019, Kasper2019}. They are characterized by a large variation (jump-like, up to a reversal) of the magnetic field, accompanied by an Alfvénic spike in the plasma velocity, together with nearly constant field strength and conserved suprathermal electron pitch angle distribution. The boundaries of the switchbacks are generally classified as Alfvénic plasma discontinuities, although not unambiguously identified as either RDs or tangential discontinuities (TDs) \citep{Larosa2021, AkhavanTafti2022, Bizien2023}. Recent studies have highlighted their importance for particle dynamics. \cite{Louarn2021} discuss the correlation between the direction of magnetic flux tubes and the development of the proton beam and find a significant beam enhancement in the vicinity of discontinuities. \cite{Shen2024} propose that discontinuities at 1 au play a unique role in enhancing plasma anisotropy by comparing theory-informed plasma anisotropy with observed electron anisotropy. 

To investigate in detail the effect of a switchback boundary (modeled as an RD) on proton kinetics in the heliosphere, we revisit a hybrid simulation of a single RD and discuss its kinetic implications, especially in generating proton beams. The remainder of this article is organized as follows. In Section 2, we introduce the simulation settings. In Section 3, we present the main simulation results and analysis from the perspectives of wave activity, plasma instabilities, and proton kinetics. Relevant discussions are presented in Section 4. 

\section{Model}

\begin{figure}
    \centering
    \includegraphics[width=0.6\linewidth]{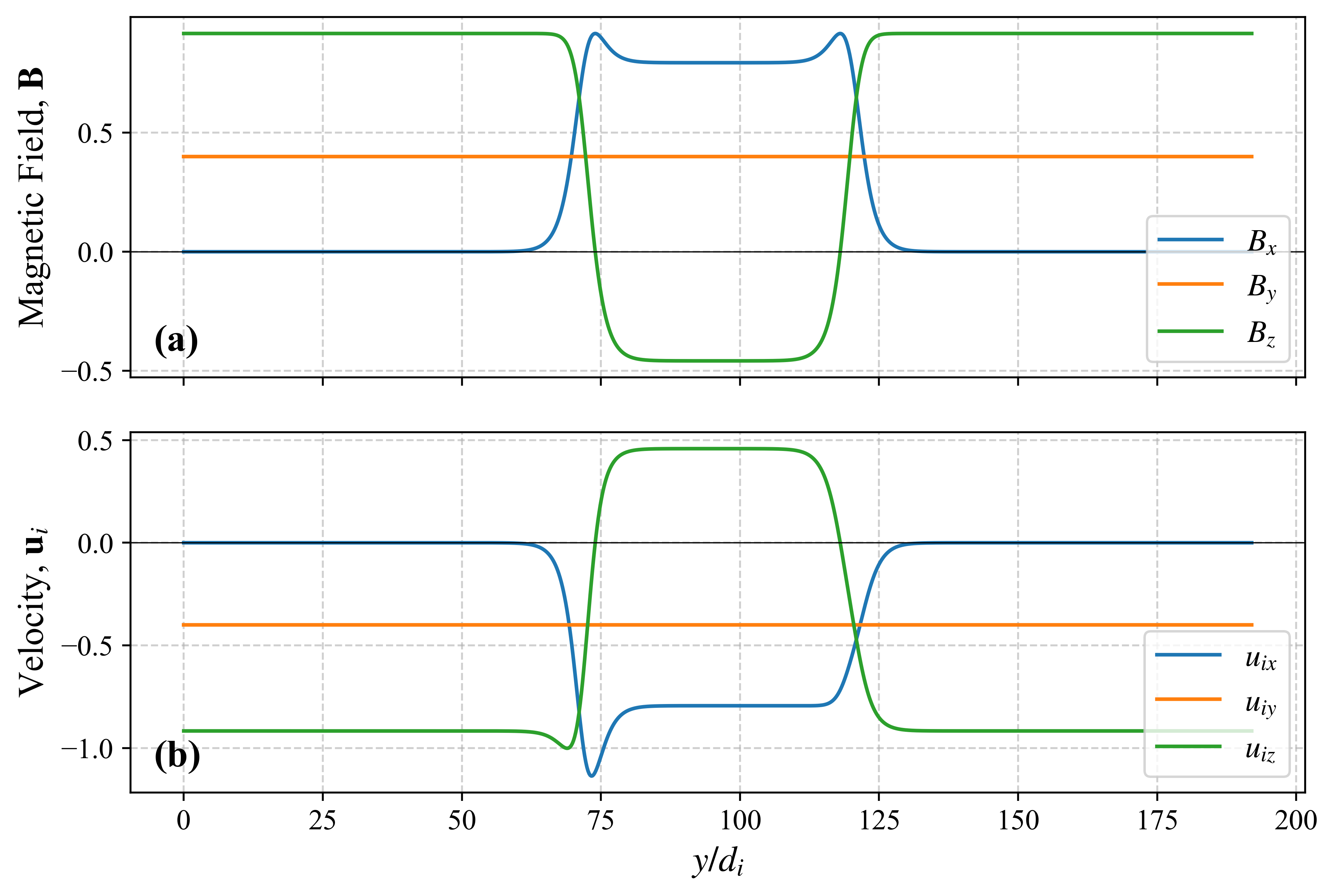}
    \caption{The initial settings of the magnetic field and the proton bulk velocity. The variables are in code units.}
    \label{fig1: initial state}
\end{figure}

The hybrid Particle-In-Cell (PIC) method treats protons as particles with mass $m_i$ and electrons as a massless fluid, under the quasi-neutrality assumption, i.e., $n_e = n_i = n$. In this work, we employ the hybrid-kinetic version of the VPIC software (referred as Hybrid-VPIC herein) \citep{Le2023Hybrid}. The simulation is initialized as follows. We use Cartesian coordinates to construct a 2D system in the $xOy$ plane, where $x\in[0,L_x)$ and $y\in[0,L_y)$. $L_x$ and $L_y$ are the lengths of the box along $x$ and $y$, respectively. The $z$-axis points out of the simulation plane. Each boundary of the simulation box is periodic. 

Inspired by frequent observations in the inner heliosphere, we initialize a pair of Alfvénic RDs characterized by rotation in the magnetic field and Alfvénically correlated plasma velocity. We define the rotation angle $\theta$, with $y$ as the direction normal to the discontinuity,
\begin{equation}
    \theta = \frac{\theta_{\max}}{2}\left(\tanh\left( \frac{y - y_1}{\Delta }\right) + \tanh\left( \frac{y_2-y}{\Delta }\right) \right),
\end{equation}
where $\theta_{\max} = 120^\circ$ is the maximum rotation angle, $y_1=72 d_i$ and $y_2 = 120 d_i$ refer to the locations of two RDs, and $\Delta =3.6d_i$ is the transition scale, where $d_i$ represents the ion skin depth.
The magnetic field is specified as 
\begin{equation} 
\left\{
\begin{aligned}
B_x &= B_{t} \sin\theta\\
B_y &= B_n\\
B_z &= B_{t} \cos\theta,
\end{aligned}
\right.
\end{equation}
where we establish a normalized constant magnitude of the magnetic field $B_0=1.0$. The constants $B_n=0.4$ and $B_t=\sqrt{B_0^2-B_n^2}$ refer to the magnetic field components normal and transverse to the discontinuity. By definition, the RD at $y_1$ is ion-sense (left-handed), meaning that as one crosses from upstream to downstream, the magnetic field rotates in the same direction as the ion gyration \citep{Richter1989}. Conversely, the RD at $y_2$ is electron-sense (right-handed), where the magnetic field rotates in the same direction as the electron gyration. The plasma density $n_0$ is uniform in the simulation domain, normalized to 1.0. In accordance with the Rankine--Hugoniot (RH) conditions of an RD, the proton bulk velocity component normal to the discontinuity $u_n=u_{iy}$ is uniformly prescribed as $u_n = - V_{A,0} B_n / B_0$, where $V_{A,0} \equiv B_0/\sqrt{\mu_0 n_0 m_p }$ is the initial Alfvén speed. As a result, the simulation is in the reference frame moving together with the discontinuity. Notably, the normalization units in the Hybrid-VPIC code set the initial $B_0$ and $V_{A,0}$ to identical values. Such a realization results in an RD static in space, while plasma flows across it. 

The simulation box is defined with dimensions $L_x=24 d_i$ and $L_y=192 d_i$, with the large extent of the y coordinate making sure that the periodic boundary condition in the y dimension (e.g., the plasma flow exiting one boundary and entering the opposite boundary) does not artificially influence the dynamic evolution of the RD pairs in the numerical domain during the time of interest.  The computational grid has $N_x\times N_y = 120\times 960$. The initial number of macro particles per cell is 2048. The time step $\Delta t$ is fixed at 0.01 $\Omega_{ci}^{-1}$, where $\Omega_{ci}$ denotes the ion (proton) cyclotron frequency. 
The governing equation for electric fields in Hybrid-VPIC is the generalized Ohm's law, 
\begin{equation}
    \label{Ohms}
    \bb{E} = -\bb{u}_{i}\times \bb{B} + \frac{1}{en_e}\bb{J}\times \bb{B} - \frac{1}{en_e} \grad p_e + \bb{R}_{ie},
\end{equation}
where the current is defined by $\bb{J}=en_e(\bb{u}_i-\bb{u}_e)$, $p_e$ is the electron pressure and $\bb{R}_{ie}=\eta \bb{J} + \eta_H\grad^2\bb{J}$ is the resistivity term. The resistivity $\eta$ and hyper-resistivity $\eta_H$ contribute to simulation stability by damping oscillations, with $\eta_H$ being particularly effective at suppressing small-scale numerical noise. Here we set $\eta=0.005B_0/ne$ and $\eta_H=0.001B_0d_i^2/ne$. By varying these two parameters and repeating the simulation, we confirm that the key physical features of the simulation are not sensitive to variations in these parameters within a reasonable range.
Assuming a vanishing initial electric field, where the electron pressure term $\frac{1}{en_e} \grad p_e$ is uniformly zero and the resistivity term $\bb{R}_{ie}$ is negligible, the initialization of the transverse component of the proton bulk velocity takes the form \citep{KraussVarban1993}:
\begin{equation}
\label{init_bulkflow}
\left\{
\begin{aligned}
    {u}_{ix} &= \frac{{u}_{iy}}{B_y} B_z - \frac{1}{en_e\mu_0}\frac{\partial B_x}{\partial y}\\
    {u}_{iz} &= \frac{{u}_{iy}}{B_y} B_x + \frac{1}{en_e\mu_0}\frac{\partial B_z}{\partial y} .
    \end{aligned}
\right.
\end{equation}
The initialization of the protons satisfies a drifting Maxwellian distribution with the bulk velocity above. The proton temperature of the initial Maxwellian distribution is isotropic. The initial proton plasma beta $\beta_{i,0} = 2 \mu_0 n_{0}\kappa T_{i,0} / B_0^2$ is set to 0.5, and the temperature ratio $T_i / T_e$ is 2.0, typical for the interplanetary plasma. The equation of state for the electrons is assumed to be adiabatic, given by $p_e n_e^{-5/3}=\rm{const}$. Fig~\ref{fig1: initial state} shows the initialization of the magnetic field and the proton bulk velocity where the profiles of $u_{ix}$ and $u_{iz}$ exhibit different trends at two RDs due to the Hall term in Equation~\eqref{init_bulkflow}. Since the rotation is only a function of $y$, all other related variables are also solely functions of $y$. 

While the setup described above is an MHD equilibrium, it is not a kinetic equilibrium and therefore we expect plasma dynamics at the kinetic level, potentially influencing microscale waves and instabilities, as we describe below.


\section{Results}

\begin{figure}
    \centering
    \includegraphics[width=1\linewidth]{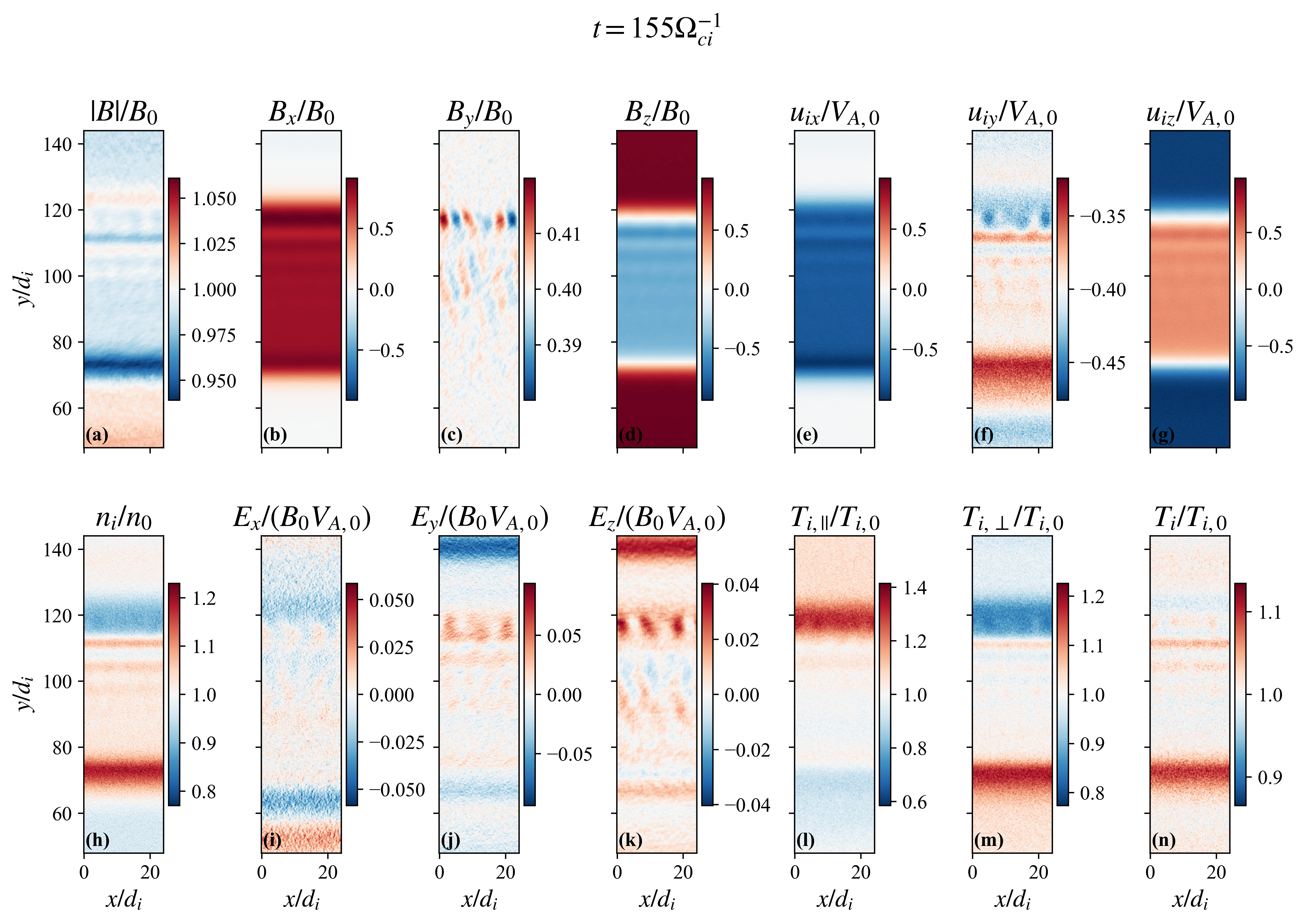}
    \caption{Key variables in the simulation. (a) The magnitude of magnetic field, (b)-(d) the three components of the magnetic field, (e)-(g) the three components of the ion bulk velocity, (h) the proton number density, (i)-(k) the three components of the electric field, (l)-(n) parallel, perpendicular and isotropic ion beta.}
    \label{fig2: simulation snapshot}
\end{figure}

\begin{figure}
    \centering
    \includegraphics[width=0.8\linewidth]{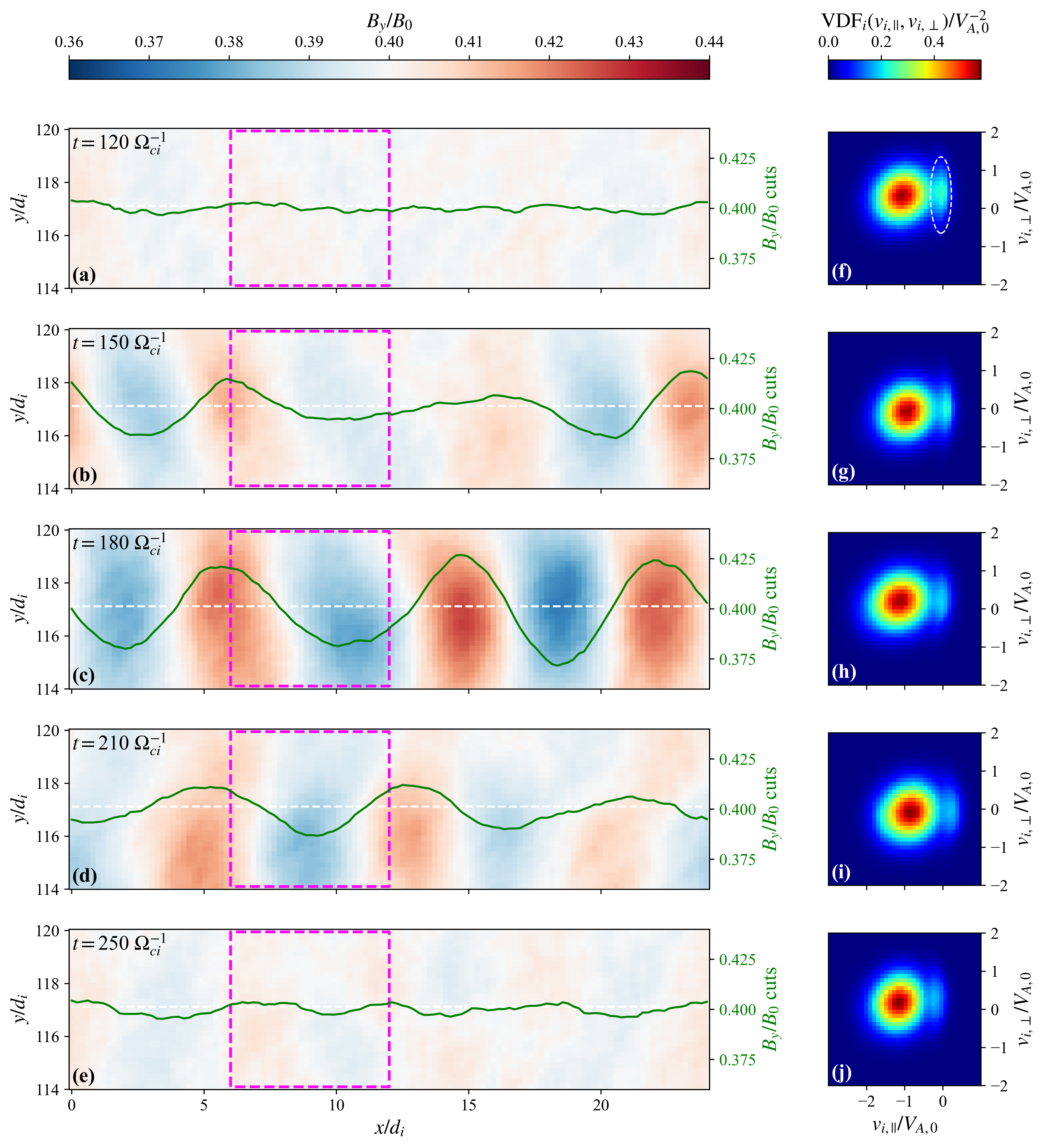}
    \caption{Excitation, growth, and damping of the left-hand polarized ICWs due to wave-particle interactions in the presence of a proton beam. (a)-(e): Spatial distribution of $B_y$ at $114d_i<y<120d_i$ where the wave activity dominates, and the green lines are cuts of $B_y$ sampled along the white dashed line. (f)-(j): Proton VDFs sampled within $114.0 d_i<y<120.0d_i$ and $6.0d_i<x<12.0d_i$ (marked with magenta dashed squares). The white dashed line marks roughly the location of the ``bean'' population in panel(f) as an example. Two panels in the same row correspond to the same simulation time. An animation of the left column, i.e. panels (a)-(e), is available. The video runs for approximately 6 seconds and covers the simulation time interval from $t=120 \Omega_{ci}^{-1}$ to $250 \Omega_{ci}^{-1}$. It demonstrates the continuous leftward propagation of the waves, clearly illustrating the dynamic process of wave amplitude growth and subsequent damping that is represented by the snapshots in the static figure.}
    \label{fig3: waveFieldAndVDF}
\end{figure}

\begin{figure}
    \centering
    \includegraphics[width=0.7\linewidth]{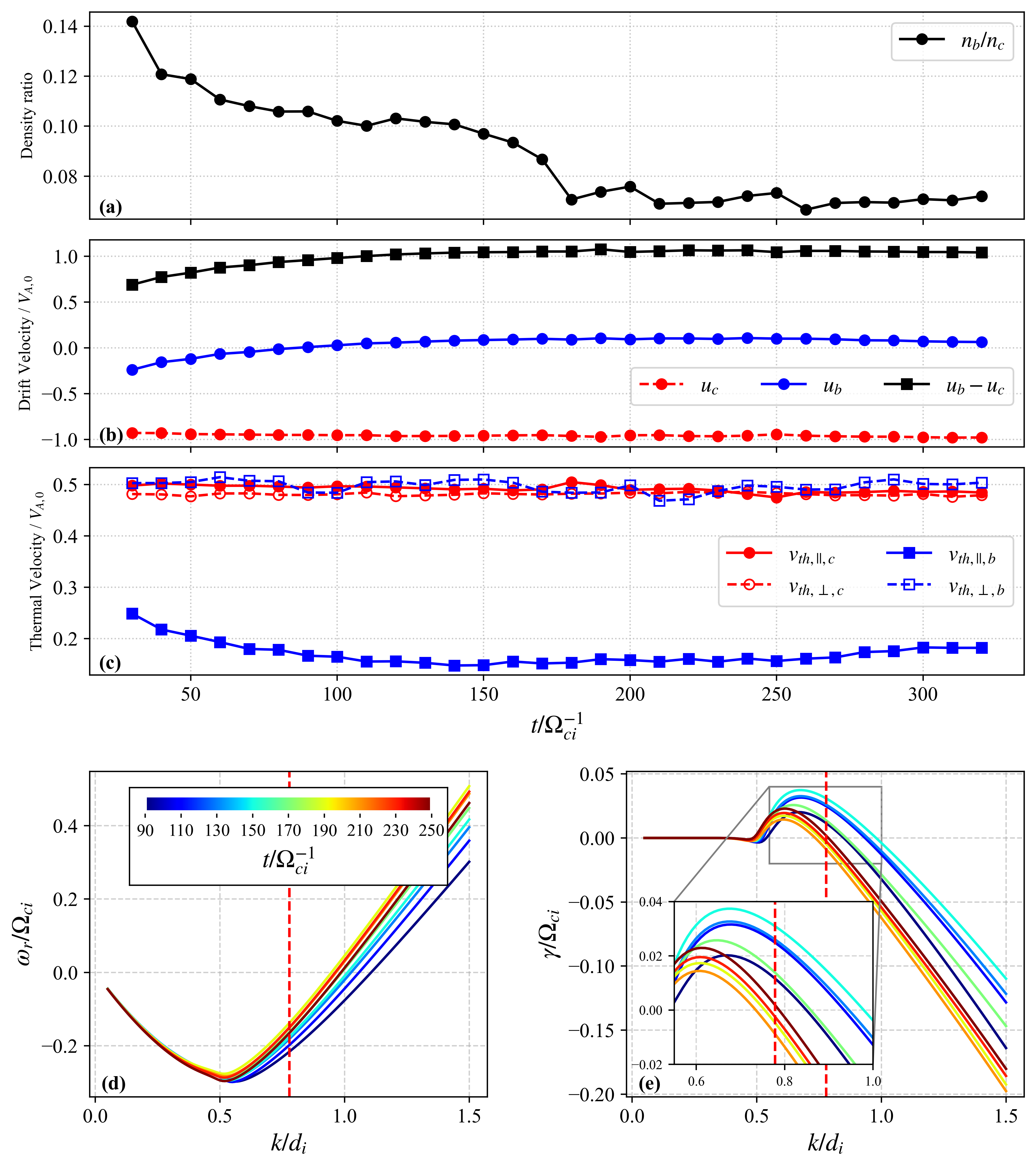}
    \caption{(a) The ratio of the fitted beam number density to the fitted core number density, as a function of time. (b) Red dashed line and red dots: fitted drift velocity of the proton core; blue solid line and blue dots: fitted drift velocity of the proton beam; black solid line and black squares: the difference between the fitted beam drift velocity and the fitted core drift velocity. (c) Red solid line and red dots: fitted parallel thermal speed of the proton core; red dashed line and red circles: fitted perpendicular thermal speed of the proton core; blue solid line and blue solid squares: fitted parallel speed of the proton beam; blue dashed line and blue hollow squared: fitted perpendicular speed of the proton beam. (d) The real frequency, as a function of wavenumber, of the wave branches with maximum growth rates at different times (marked with the colorbar). (e) The imaginary frequency, as a function of the wavenumber, of the same wave branches in (d). The red dashed line in (d) and (e) refers to the estimated wavenumber in the simulation.}
    \label{fig4: fittingRes}
\end{figure}

\subsection{Overview of Simulation Results for a Pair of Kinetic RDs}
The simulation is run until $t = 500\Omega_{ci}^{-1}$. During the whole simulation, the two RDs are static in space as expected. No significant broadening occurs. 
To highlight key dynamic features, we present a representative snapshot of the simulation at $t = 155 \Omega_{ci}^{-1}$ in Figure~\ref{fig2: simulation snapshot}, where the buffer regions (i.e., $y < 48d_i$ and $y > 144d_i$) have been cropped to focus on the regions of interest. In this figure, we display the magnetic field strength and components, ion bulk velocity components, electric field components, and proton number density. Additionally, we define three key ion temperature quantities, isotropic temperature ($T_i$), parallel temperature ($T_{i\parallel}$), and perpendicular temperature ($T_{i\perp}$) with respect to the background magnetic field, using the ion thermal pressure tensor $\mathsf{P}_i$. These are given by:
\begin{equation}
\left\{
\begin{aligned}
    T_i &= p_{i} /n_i\\
    T_{i||} &= \mathsf{P}_i\bdbldot\hat{\bb{b}}\hat{\bb{b}}/n_i \\
    T_{i\perp} &= \frac{1}{2}\left(3p_{i}-\mathsf{P}_i\bdbldot\hat{\bb{b}}\hat{\bb{b}}\right)/n_i,
    \end{aligned}
\right.
\end{equation}
where $p_{i}=\mathrm{Tr}(\mathsf{P}_i)/3$ is the isotropic ion thermal pressure, and $\hat{\bb{b}}=\bb{B}/B$. The Boltzmann constant is 1 in simulation normalized units. 

The most interesting features in Figure~\ref{fig2: simulation snapshot} are wave-like activities around the upper RD with electron-sense (right-handed) rotation of the magnetic field fluctuations, especially pronounced in $B_y$. Fluctuations are concentrated within the downstream portion of the RD transition that spans approximately 5 $d_i$ in the $y$ direction. There are also fluctuations that extend slightly away from the upper RD on the downstream side, but with smaller amplitude. From here on, we focus on the analysis of the regions with the strongest fluctuations. Correlated fluctuations of $u_{iy}$, $E_y$, and $E_z$ are also visible, but not significantly observed in $|B|$ and $n_i$, suggesting that the fluctuations are not compressible. 

\subsection{Wave Activity and Associated VDF Evolution}
We have conducted further quantitative analysis on the fluctuations. 
The left column of Figure~\ref{fig3: waveFieldAndVDF} illustrates the evolution of $B_y$ in the transition region of interest. An animated version of this column is available online, showing the continuous temporal evolution of the wave fluctuations from $t=120 \Omega_{ci}^{-1}$ to $250 \Omega_{ci}^{-1}$. Figure~\ref{fig3: waveFieldAndVDF}a-c depict the growth phase: slight fluctuations (Fig.~\ref{fig3: waveFieldAndVDF}a) develop into moderate waves (Fig.~\ref{fig3: waveFieldAndVDF}b) and reach maximum amplitude in Figure~\ref{fig3: waveFieldAndVDF}c. Subsequently, Figure~\ref{fig3: waveFieldAndVDF}c-e show the damping process: the amplitude of fluctuations decreases and eventually returns to the initial low level (Fig.~\ref{fig3: waveFieldAndVDF}e). We concluded that these fluctuations are left-hand (LH) circularly polarized waves propagating antiparallel to the background magnetic field. The compressible component of the waves has an amplitude less than $1\%$ of the transverse component, so these waves are considered incompressible. Although these fluctuations experience growth and damping over the course of the simulation, we focus here on reporting a set of representative wave characteristics concerning the moment they are fully developed, when wave amplitudes and coherence are strongest. At this stage, the waves exhibit a representative wavelength $\lambda_{rep}=8.07d_i$, a frequency of $0.137\Omega_{ci}$, and a corresponding phase speed $V_{p,rep}=0.302 V_{A,0}$ (measured in the plasma rest frame). These quantities vary by less than 5 percent when the wave activity is clearly observable. The key characteristics mentioned above suggest that the most probable wave mode is the ion cyclotron wave (ICW), as we will discuss in more detail later on. 

The right column of Figure~\ref{fig3: waveFieldAndVDF} presents the proton VDFs sampled at the same time of their corresponding wave fields. For example, Panels (a) and (f) suggest that before the waves grow, there is a beam-like secondary population with a velocity difference relative to the bulk of the plasma population along the background magnetic field (e.g. the population marked with the white dashed line in Figure~\ref{fig3: waveFieldAndVDF}f). For simplicity, we name this main population and the beam population as the ``core'' and the ``beam'' of the VDF, respectively. The beam elongates in the direction perpendicular to the background magnetic field. The relative abundance of the beam population evolves with the wave lifecycle. When the waves start to grow, the beam population appears to comprise a smaller percentage of the total number of particles. After the waves are damped, the beam appears to constitute a minimum percentage. We observe a trend where, as the waves grow, protons lose bulk kinetic energy. Conversely, when the waves are damped, part of this energy is transferred back to the protons' kinetic energy through wave-particle interaction. 

In Figure~\ref{fig4: fittingRes} we present the results from the application of a two-component fitting algorithm to further investigate VDFs and related wave activities. The fitting assumes each VDF to consist of two components, the proton core (denoted by ``c'') and the proton beam (denoted by ``b''). The two components are both approximated as bi-Maxwellian, each characterized by (potentially different) parallel and perpendicular thermal speeds relative to the background magnetic field, as 
\begin{equation}
f(v_{||}, v_{\perp}) = \frac{n}{ (2\pi)^{3/2} v_{th,||} v_{th,\perp}^2 } \exp \left( - \frac{(v_{||} - u)^2}{2 v_{th,||}^2} - \frac{v_{\perp}^2}{2 v_{th,\perp}^2} \right).
\end{equation}
The plots indicate that during the early phase of the simulation (until $t\sim 100\Omega_{ci}^{-1}$, when significant wave growth commences), the beam evolves to converge on $n_b/n_c\sim 0.1$, $u_{b}-u_{c} \sim V_{A,0}$ and $v_{th,||,b}\sim 0.15 V_{A,0}$. Subsequently, the characteristics of the beam remain nearly constant until the amplitude reaches its maximum around $t=150\Omega_{ci}^{-1}$, prior to a significant drop in $n_b/n_c$. After the waves are damped, $n_b/n_c$ reaches a minimum percentage, exhibiting slight fluctuations. Beyond $t\sim 100\Omega_{ci}^{-1}$, the drift velocities and the thermal velocities of both populations appear to converge to their asymptotic values. In the whole time range presented, the beam has a significant temperature anisotropy, with $T_\perp/T_\|\gtrsim4$.

We use a dispersion solver, ``Plasma Kinetics Unified Eigenmode Solutions'' (PKUES, \cite{Luo2022}) for linear plasma systems to obtain the dispersion relation and growth rates of probable waves based on the fitted parameters obtained above. Panels (d) and (e) of Figure~\ref{fig4: fittingRes} present the key results of applying PKUES to the simulated plasma conditions. At the wavenumber estimated from the simulation $k_{rep}=2\pi/\lambda_{rep}$, the waves are predicted to have a phase speed that matches the estimated representative value, $V_{p,linear}\sim V_{p,rep}$. 
The sign of the real part of wave frequency, $\omega_r(k_{rep})$, is negative, which is consistent with their anti-parallel propagation. Combining $\omega_r(k_{rep})<0$ and the fluctuating electromagnetic fields given by PKUES (not shown here), the waves are LH circularly polarized, thereby further supporting the idea that the wave mode as ICW. On the other hand, the peak of the curve does not coincide with $k_{rep}$, indicating that the wavevector observed in the simulation differs from the one associated with the maximum growth rate. This discrepancy can be attributed to the confinement of waves within a specific spatial range in $y$ and the influence of the periodic boundary condition. The growth rate at $k_{rep}$ predicts potential wave behavior in the simulation. The growth rate at $t=110$, $130$, and $150\Omega_{ci}^{-1}$ during the active growth phase is much larger than 0, while subsequently from $t=150\Omega_{ci}^{-1}$ the growth rate decreases and finally becomes negative, which also aligns with the simulation results. Therefore, the growing and damping process can be predicted by linear wave theory. In conclusion, as the sole input to PKUES, the proton beam is the primary cause of the observed wave activity.

\subsection{Origin of the Proton Beam Responsible for the Wave Activity}
\begin{figure}
    \centering
    \includegraphics[width=1\linewidth]{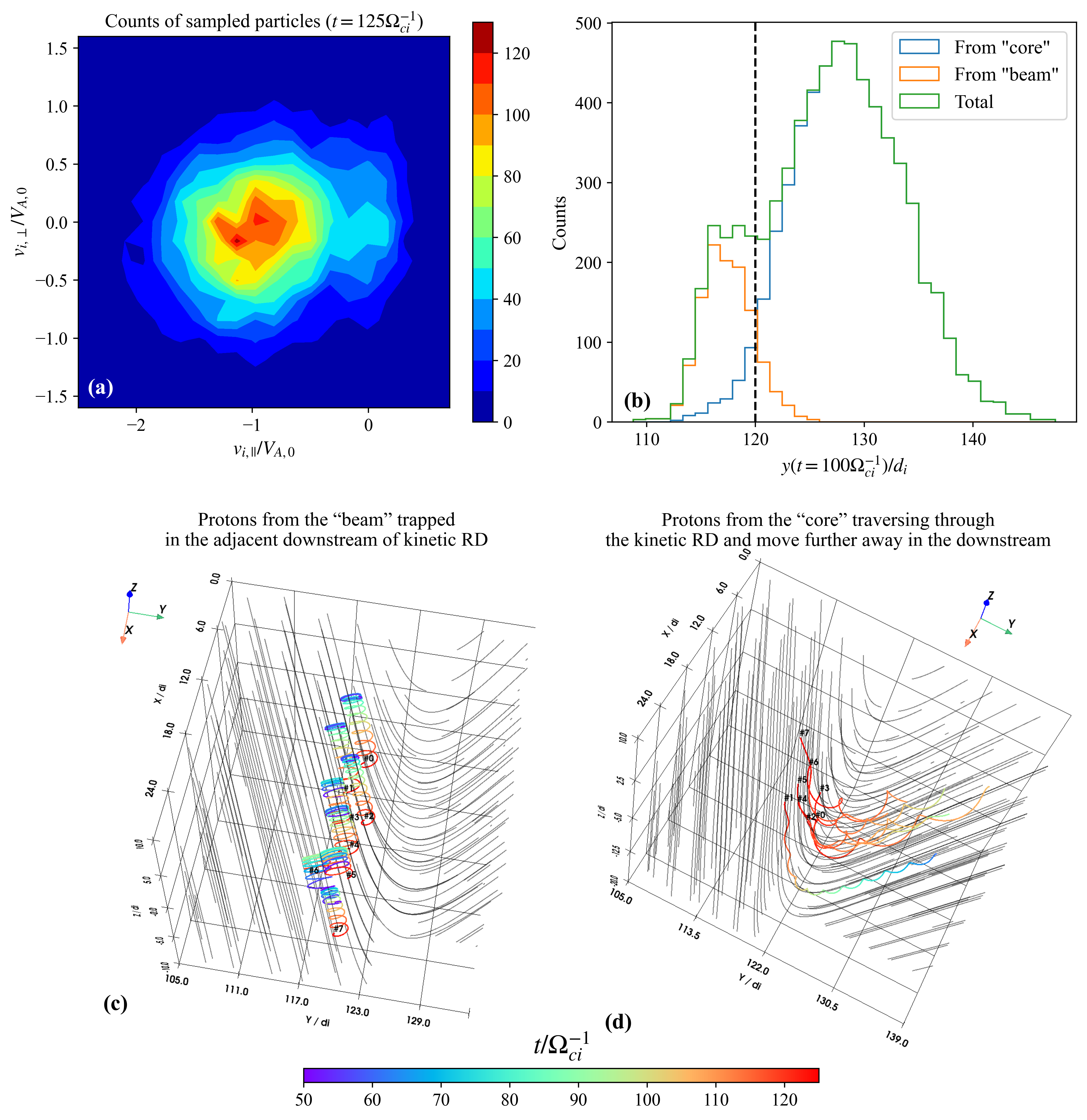}
    \caption{(a) Velocity distribution of downsampled traced particles at $t=125\Omega_{ci}^{-1}$. These particles are sampled from the same region as in Figure~\ref{fig3: waveFieldAndVDF}f-j. 
    (b) Distribution of $y$-positions of traced particles at $t=100\Omega_{ci}^{-1}$. The black vertical dashed line corresponds to the location of the RD of interest.  
    (c)-(d) Sampled particle trajectories from the ``beam'' and ``core'' groups. The colors on trajectories indicate time. The black solid lines show the local magnetic field geometry.
    }
    \label{fig5: tracing}
\end{figure}

A critical question remains: how does the proton beam originate? To address this question, we trace selected ensembles of particles in the simulation. 
The traced particles constitute a downsampled subset of all particles within the region of interest where the VDF in Figure~\ref{fig3: waveFieldAndVDF} is plotted, corresponding to $t=125\Omega_{ci}^{-1}$ during the early stage of the wave growth phase. The downsampling factor with respect to the total number of particle within the region of interest is 256. The velocity distribution of the sampled particles is shown in Figure~\ref{fig5: tracing}a. The proton beam is weakened but still recognizable. 

By simply splitting traced particles into two groups with the criterion $v_{i,||}>-0.3 V_{A,0}$ for the ``beam'' and the complementary criterion for the ``core'' as marked in Figure~\ref{fig5: tracing}b, the histogram shows that at $t=100\Omega_{ci}^{-1}$, i.e. approximately 4 proton gyro-periods $T_{ci} = 2\pi / \Omega_{ci}^{-1}$ before the VDF is sampled, the core particles and the beam particles exhibit distinct distribution in $y$. The core particles move along with the bulk proton flow upstream of the RD, while the beam particles appear to persistently remain on the downstream part of the RD transition. Although $v_{i,||}>-0.3 V_{A,0}$ is an approximate criterion, two distinct peaks are indeed present in the $y|_{t=100 \Omega_{ci}^{-1}}$ distribution of all sampled particles, and this criterion effectively distinguishes between them.
Eight typical particle trajectories selected from the ``beam'' group, shown in Figure~\ref{fig5: tracing}c, suggest that these particles experience a reflection process that resembles magnetic mirroring. To visualize the particle cyclotron motion in 3D space, we integrated each particle's $v_z$ to obtain its $z$-coordinate. These particles are traced from $t=50\Omega_{ci}^{-1}$ until $t=125\Omega_{ci}^{-1}$ with trajectory colors indicating time. The $x$-coordinates of all particles are shifted to separate the trajectories, without affecting the relationship between the trajectories and the magnetic field lines, since the field lines exhibit translational symmetry along the $x$-axis, given the moderate wave amplitude over the background magnetic field. These traced particles experience reflection at varying times (indicated by colors), suggesting that reflection occurs continuously. Several particle trajectories from the ``core'' are also shown in Figure~\ref{fig5: tracing}d for comparison. These trajectories demonstrate that proton core population traverses the magnetic field rotation region, with no evidence of magnetic reflection, which is distinct from the bouncing behavior of beam particles.

\begin{figure}
    \centering
    \includegraphics[width=1\linewidth]{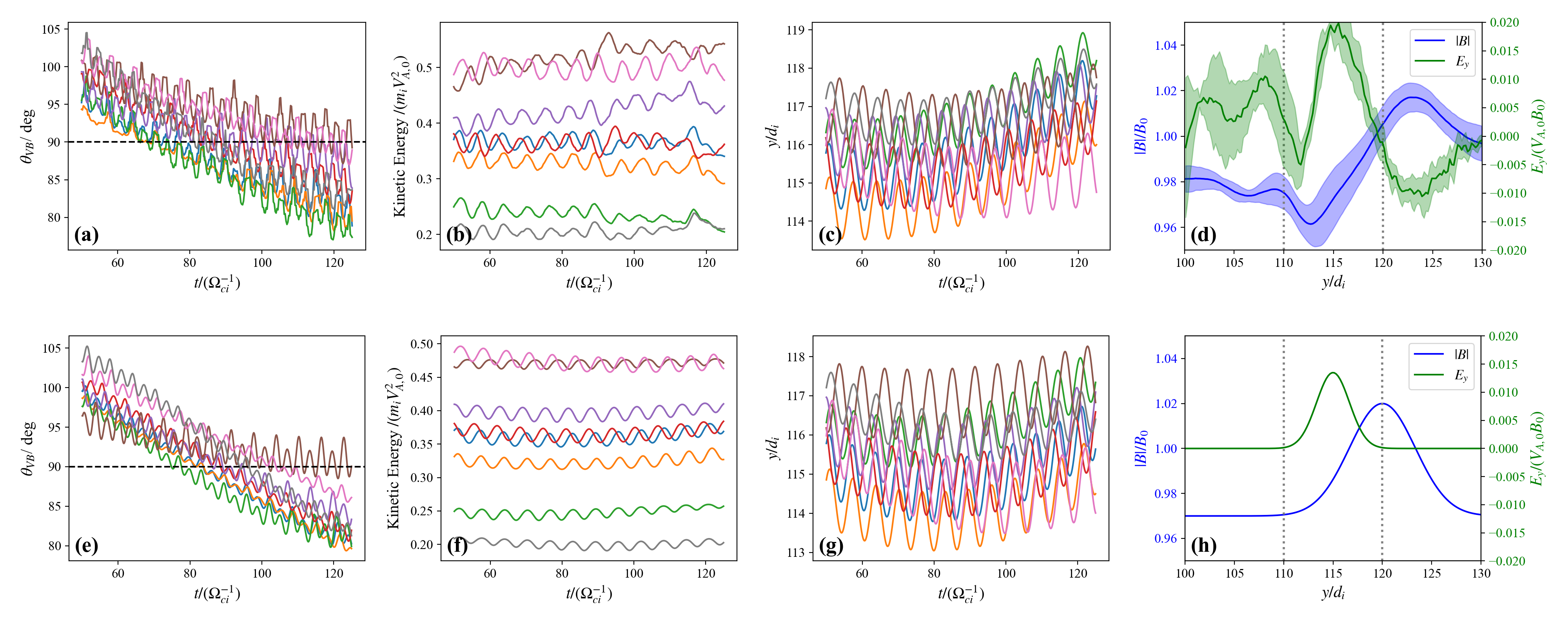}
    \caption{The top row is from the hybrid PIC simulation. (a) pitch angle, with a black dashed line at $90^\circ$ indicating the reflection threshold, (b) kinetic energy, and (c) $y$-coordinate of the same eight traced particles as in Figure~\ref{fig5: tracing}c. (d)Solid lines: $y$-component of the electric field ($E_y$, green) and magnitude of the magnetic field ($|B|$, blue) averaged over time from $t=50\Omega_{ci}^{-1}$ to $t=125\Omega_{ci}^{-1}$, as a function of $y$ in the region of interest. Shaded regions: the temporal standard deviation of the corresponding variables.} Two black dashed lines mark a region in between where reflections mainly happen.The bottom row is the counterpart of the top row from the test-particle run in a preset electromagnetic field environment that resembles the simulated situation.
    \label{fig6: pitchAngleTest}
\end{figure}

A gradient in magnetic field strength is a common trapping factor in magnetic mirror structures and magnetic holes \citep{Post1959,Balikhin2009,Arro2024}. However, the simulated RD is initialized without a gradient in magnetic field strength. To identify the key mechanism of trapping, we further examine the relationship between proton kinetics and local electromagnetic field by presenting key diagnostics in Figure~\ref{fig6: pitchAngleTest}. The top row presents pitch angles, kinetic energy, and $y$-coordinate over time for eight typical particles sampled above, alongside the local electromagnetic field. The pitch angles change from $\sim 100$ degrees to $\sim80$ degrees and simultaneously fluctuate periodically with a frequency slightly larger the proton gyro-frequency. One can compare the frequency to the oscillation frequency in the $y$-coordinate in panel (c), which is approximately equal to $\Omega_{ci}$. The fluctuation of pitch angles is attributed to the kinetic-scale RD: while particles are gyrating, they also encounter a rotation in the local magnetic field; the crossing of this rotation occurs on the same timescales as their gyration. Moreover, $|B|$ consistently exhibits a gradient along the $+y$ direction, corresponding to a mirror force directed along $-y$, which opposes particle reflection. Therefore, the reflection is not due to the magnetic gradient. The actual cause of the reflection is identified as the static electric field in the RD reference frame, especially the persistent positive enhancement of $E_y$ around the region of interest, as shown in panel (d).

To verify this idea, we conducted a test-particle simulation using the standard Boris algorithm \citep{boris1970relativistic}, incorporating three key factors: a formulated magnetic field rotation as the initial state; a Gaussian profile resembling the electric field ``bump'' with comparable potential; and a gradient of magnetic field strength derived from a Gaussian curve, as depicted in Figure~\ref{fig6: pitchAngleTest}h. The resulting pitch angle, kinetic energy, and $y$-coordinate of the test particles are shown in the bottom row of Figure~\ref{fig6: pitchAngleTest} and closely resemble those observed in the simulation. We note that if we remove any one factor, this similarity is no longer maintained.

\begin{figure}
    \centering
    \includegraphics[width=1.0\linewidth]{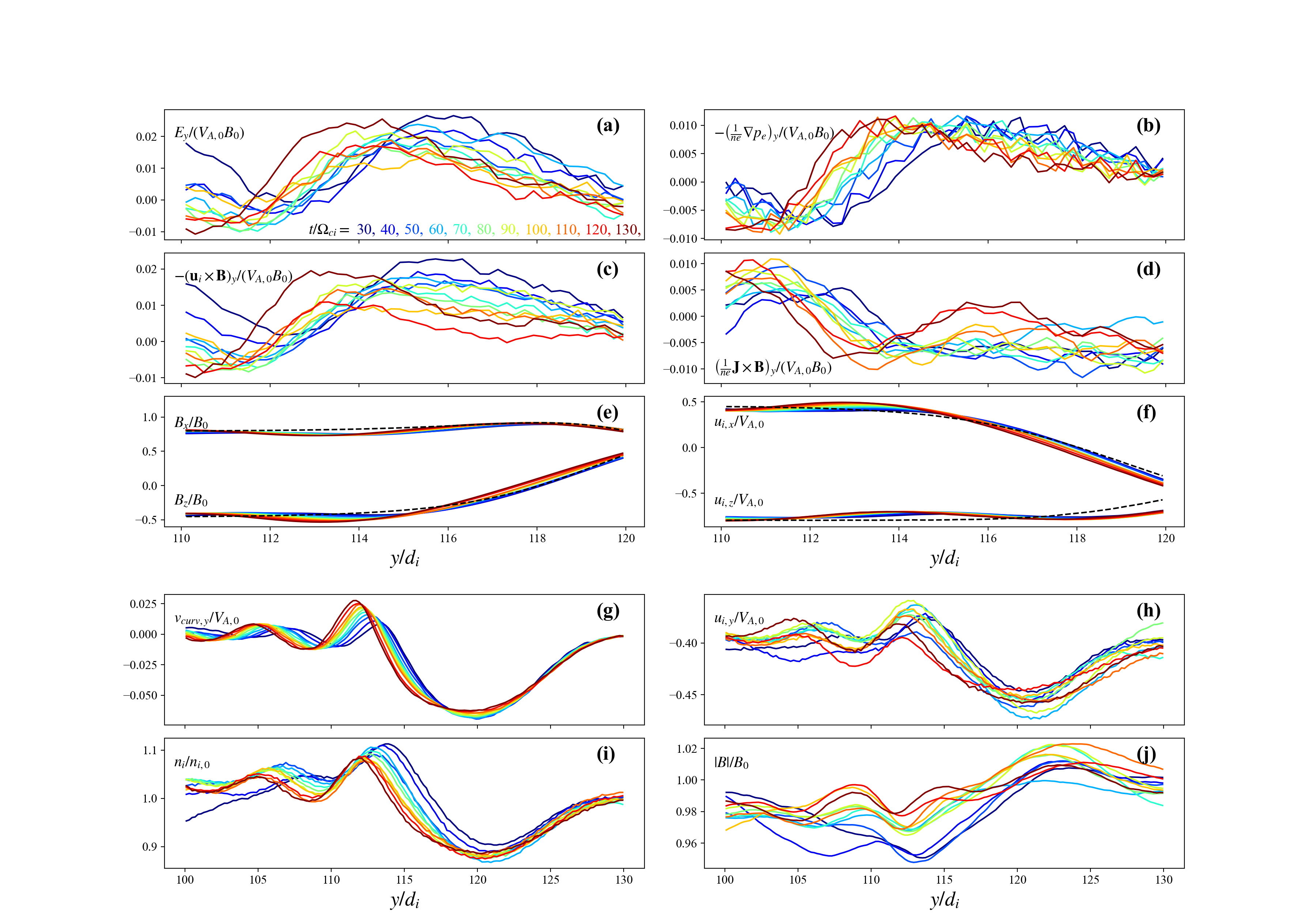}
    \caption{Panels (a)-(e) show $E_y$, the electron pressure term, the convective electric field, the Hall electric field, and the resistivity term at the regions of interest versus time, respectively. Panels (f)-(h) show the $y$-component of the curvature drift velocity (${v}_{{\boldsymbol{\nabla} B}, y}$), the ion density and magnetic field strength. (i) $B_x$ (from upper left to lower right) and $B_z$ (from lower left to upper right) versus time. (j) $u_{i,z}$ (from upper left to lower right) and $u_{i,x}$ (from lower left to upper right) versus time. Colors denote the simulation time. The black dashed lines in panel (i) and (j) denote the initial distributions of $B_x$, $B_z$, $u_{i,z}$ and $u_{i,x}$, to to illustrate their deviation during evolution.}
    \label{fig7:discussionOfBandE}
\end{figure}

A further brief analysis can help us understand the measured electromagnetic fields at the kinetic scales. 
Figure~\ref{fig7:discussionOfBandE}a-f present decompositions of $E_y$ versus time as in Equation~\ref{Ohms}. From $t=40\Omega_{ci}^{-1}$ to $t=120\Omega_{ci}^{-1}$, approximately when the proton beam is forming dynamically, the $E_y$ enhancement remains steady and is primarily attributed to the convective electric field $-\bb{u}_i\times \bb{B} $. The electron pressure term and the Hall term are of comparable magnitude and approximately balance each other. Note that at $t=0$ the simulation is initialized with a vanishing electric field, with the Hall term and the convective term balancing each other. We also plot the $x$- and $z$-components of the magnetic field and the bulk proton velocity to investigate their contribution to the convective electric field, see panels (i) and (j). The black dashed lines here denote the initial distributions of $B_x$, $B_z$, $u_{i,z}$ and $u_{i,x}$. The deviation from these initial states due to evolution indicates no leading term that accounts for the enhancement in $-\left(\bb{u}_i\btimes\bb{B} \right)_y$, suggesting that the structures of the current density ($\bb{J}$) around an evolving RD at kinetic scales are complicated.
Figure~\ref{fig7:discussionOfBandE}g-j present several quantities associated with the evolution of the magnetic strength. In Figure~\ref{fig7:discussionOfBandE}g we plot the $y-$component of the curvature drift velocity resulting from the magnetic field topology. The curvature drift velocity is defined as
\begin{equation}
    \bb{v}_{\mathrm{curv} B} = \frac{m v_{\parallel}^2}{q B} \bb{b} \btimes [(\bb{b}\bcdot \grad)\bb{b}].
\end{equation}
The calculated curvature drift is highly correlated with the $y-$ component of the ion bulk flow velocity shown in Figure~\ref{fig7:discussionOfBandE}h, suggesting that the curvature drift effect contributes to the deviation of $u_{iy}$ from its initial state. Furthermore, the density profile in Figure~\ref{fig7:discussionOfBandE}i is in phase with $u_{iy}$, and is anti-correlated with $|B|$ ( Figure~\ref{fig7:discussionOfBandE}j). This phase relationship resembles that of compressible waves, specifically (kinetic) slow modes propagating in the $+y$ direction (from the downstream part of the RD to the upstream) against a background bulk flow in the $-y$ direction. In summary, the variation in $|B|$ is governed by a compressible (kinetic) slow mode propagating upstream at the RD interface, where the initial fluctuations are driven by curvature drift.

\section{Summary and Discussion}

We have conducted a hybrid PIC simulation of a switchback-like boundary layer modeled as a kinetic rotational discontinuity. Our double-periodic simulation initializes a pair of kinetic RDs with a Hall velocity transition; we find that the electron-sense (right-handed) RD traps a significant population of protons, forming a proton beam population aligned with the background magnetic field that is strong enough to excite ion cyclotron waves. The key mechanism driving this beam formation is a static electric field component that is aligned with the RD normal and remains relatively static in the RD reference frame. The convective electric field is the primary contributor to this electric field. Although this term is initially zero, it develops as the simulation progresses. Furthermore, the presence of the beam makes $u_{i,x}$ closer to zero, which in return enhances the convective electric field term $-(\bb{u}\times\bb{B})_y$ aligned with the RD normal, thus creating a self-sustaining mechanism for the local electric potential. 

Three factors determine the dynamics of the beam protons: the electric potential, the magnetic field bump, and the field rotation. We conducted test particle simulations with magnetic and electric field profiles mimicking those we find in the hybrid PIC run when wave activity is detected; these test-particle simulations indicate that the electric potential is the dominant cause of trapping. The magnetic field modulates the trapping process, and its slight enhancement also prevents ensembles of particles from going further to the upstream. The rotation direction does not have a direct and significant effect on the proton kinetics. When reversing the rotation while keeping all other factors the same, our simulations show that fewer than 20\% protons escape the trap. This might result from the electron-sense (right-handed) of rotation, which means that passing protons experience the background changes in an opposite direction to their gyration. 
As a result, proton pitch angles exhibit larger fluctuations, making it easier for them to reach the $90^\circ$ threshold. However, it is non-straightforward to isolate this effect in the hybrid simulation. When the rotation is reversed, the local variations in the magnetic field also change. This, in turn, alters the curvature drift and modifies the balance of currents will be different, so the same static electric field and magnetic bump do not form in the downstream portion of the transition. That is the reason why we observe the proton beam and wave-like activity in the vicinity of the upper RD at $y=120 d_i$ only in the electron-sense (right-handed) rotation, whereas no trapping occurs at the lower ion-sense (left-handed) RD at $y\sim 72d_i$. We also confirm that in the simulation, there is no such electric field around $y\sim 72d_i$, even though there is a gradient of magnetic field strength with a corresponding gradient force towards $+y$ in the vicinity of the lower RD. These features highlight the importance of the electric field in trapping protons.

As shown in Figure~\ref{fig4: fittingRes}, the proton beam has a lower parallel temperature and a comparable perpendicular temperature to the core population. Its drift velocity is very close to the Alfvén speed. These features do not match well with the observations of \cite{Louarn2021}, which reported a higher parallel temperature and a lower perpendicular expansion. In the context of switchbacks in the inner heliosphere, whether such a beam undergoes other processes around these discontinuities is still an open question. It is also instructive to briefly compare the beam with the ``hammerhead'' population \citep{Verniero2022}. While similar in temperature anisotropy, they differ significantly in drift speed and background magnetic topology. The hammerhead population is typically observed along with background radial magnetic field without sharp field rotations, and exhibits a much higher drift speed (up to $5-6 V_A$). \cite{Malara2021} also discussed the trapping mechanisms of RDs, but for energetic particles instead of the warm solar wind plasma investigated here. Nevertheless, understanding the magnetic topology of RDs and their kinetic effects on plasma remains a valuable scientific issue as we probe the finer scales of the interplanetary plasma. 

Such field-aligned proton beams generated at RD-like boundaries may not remain passive structures. Solar Orbiter observations have shown that non-field-aligned beams can efficiently excite fast magnetosonic/whistler (FM/W) waves in the solar wind \citep{Zhu2023}. Moreover, linear and nonlinear kinetic theories predict that proton beams with temperature anisotropy can further drive both FM/W and Alfvén/ion-cyclotron (A/IC) instabilities, depending on their drift speed and anisotropy \citep{xiang2021linear}. Therefore, the beams produced in our simulations could serve as seed populations for wave growth, suggesting that RD-generated beams may play an active role in plasma heating and wave–particle interactions in the solar wind.

Our simulation also emphasizes the role of the relatively static electric field in trapping particles. Relatively static electric trapping, which has been discussed in the context of dynamic structures and processes like magnetic reconnection and coherent structures \citep{Egedal2005,Hutchinson2020,Xie2024}, might also be worth investigating in observations of rotational discontinuities.

The primary limitation of this study is its two-dimensional (2D) geometry. Key 3D features of realistic switchback structures, such as the orientation of flux tubes and their angle relative to the Parker spiral, can only be fully captured in 3D simulations \citep{Laker2023}. Here, we also avoided the use of the piston method which fixes the parameters downstream of the RD as \cite{KraussVarban1995}, because reflecting plasma flows on the boundary might affect particle kinetics and instabilities in the volume. However, extending the simulation box to obtain a larger domain transit time is a viable, though imperfect method to preventing the re-mixing of plasma from the periodic boundary; this will be investigated in future work. Finally, since Alfvénic discontinuities are frequently embedded in high speed solar wind streams, where the core-beam distributions of protons are commonly observed, future simulations with an initial VDF consisting of a core and a beam are also a promising avenue for investigation.

\begin{acknowledgments}
The work at Peking University is supported by NSFC (42530105, 42241118, 42174194, 42150105, and 42204166), by National Key R\&D Program of China (2021YFA0718600 and 2022YFF0503800), and by CNSA (D010301, D010202, D050103).
F.B.\ acknowledges support from the FED-tWIN programme (profile Prf-2020-004, project ``ENERGY'') issued by BELSPO, and from the FWO Junior Research Project G020224N granted by the Research Foundation -- Flanders (FWO). 
LP\ acknowledges support from a PhD fellowship in fundamental research awarded by the Research Foundation -- Flanders (FWO), under grant number 11PCB24N.
The resources and services used in this work were provided by the VSC (Flemish Supercomputer Center), funded by the Research Foundation - Flanders (FWO) and the Flemish Government.
\end{acknowledgments}

\bibliography{references}{}

@article{he2018plasma,
  title={Plasma heating and Alfv{\'e}nic turbulence enhancement during two steps of energy conversion in magnetic reconnection exhaust region of solar wind},
  author={Jiansen, He and Xingyu, Zhu and Yajie, Chen and Chadi, Salem and Michael, Stevens and Hui, Li and Wenzhi, Ruan and Lei, Zhang and Chuanyi, Tu},
  journal={The Astrophysical Journal},
  volume={856},
  number={2},
  pages={148},
  year={2018},
  publisher={IOP Publishing}
}

@article{duan2023kinetic,
  title={Kinetic Features of Alpha Particles in a Pestchek-like Magnetic Reconnection Event in the Solar Wind Observed by Solar Orbiter},
  author={Duan, Die and He, Jiansen and Zhu, Xingyu and Zhuo, Rui and Wu, Ziqi and Nicolaou, Georgios and Huang, Jia and Verscharen, Daniel and Yang, Liu and Owen, Christopher J and others},
  journal={The Astrophysical Journal Letters},
  volume={952},
  number={1},
  pages={L11},
  year={2023},
  publisher={IOP Publishing}
}

@article{zhuo2024oblique,
  title={Oblique Compressible Waves in the Reconnection Exhaust Region Embedded in the Inner Heliospheric Current Sheet Observed by Parker Solar Probe},
  author={Zhuo, Rui and He, Jiansen and Duan, Die and Zhu, Xingyu and Hou, Chuanpeng},
  journal={The Astrophysical Journal},
  volume={969},
  number={1},
  pages={47},
  year={2024},
  publisher={IOP Publishing}
}

@article{xiang2021linear,
  title={Linear and nonlinear effects of proton temperature anisotropy on proton-beam instability in the solar wind},
  author={Xiang, L and Lee, KH and Wu, DJ and Yu, HW and Lee, LC},
  journal={The Astrophysical Journal},
  volume={916},
  number={1},
  pages={30},
  year={2021},
  publisher={IOP Publishing}
}

@article{Feldman1973,
  title = {On the origin of solar wind proton thermal anisotropy},
  volume = {78},
  ISSN = {0148-0227},
  url = {http://dx.doi.org/10.1029/JA078i028p06451},
  DOI = {10.1029/ja078i028p06451},
  number = {28},
  journal = {Journal of Geophysical Research},
  publisher = {American Geophysical Union (AGU)},
  author = {Feldman,  W. C. and Asbridge,  J. R. and Bame,  S. J. and Montgomery,  M. D.},
  year = {1973},
  month = oct,
  pages = {6451–6468}
}

@article{Marsch1982,
  title = {Solar wind protons: Three‐dimensional velocity distributions and derived plasma parameters measured between 0.3 and 1 AU},
  volume = {87},
  ISSN = {0148-0227},
  url = {http://dx.doi.org/10.1029/JA087iA01p00052},
  DOI = {10.1029/ja087ia01p00052},
  number = {A1},
  journal = {Journal of Geophysical Research: Space Physics},
  publisher = {American Geophysical Union (AGU)},
  author = {Marsch,  E. and M\"{u}hlh\"{a}user,  K.‐H. and Schwenn,  R. and Rosenbauer,  H. and Pilipp,  W. and Neubauer,  F. M.},
  year = {1982},
  month = jan,
  pages = {52–72}
}

@article{gary_electromagnetic_1991,
	title = {Electromagnetic ion/ion instabilities and their consequences in space plasmas: {A} review},
	volume = {56},
	issn = {1572-9672},
	url = {https://doi.org/10.1007/BF00196632},
	doi = {10.1007/BF00196632},
	number = {3},
	journal = {Space Science Reviews},
	author = {Gary, S. Peter},
	month = may,
	year = {1991},
	pages = {373--415},
}

@article{Hellinger2011,
  title = {Proton core-beam system in the expanding solar wind: Hybrid simulations: PROTON CORE-BEAM SYSTEM IN SOLAR WIND},
  volume = {116},
  ISSN = {0148-0227},
  url = {http://dx.doi.org/10.1029/2011JA016940},
  DOI = {10.1029/2011ja016940},
  number = {A11},
  journal = {Journal of Geophysical Research: Space Physics},
  publisher = {American Geophysical Union (AGU)},
  author = {Hellinger,  Petr and Trávníček,  Pavel M.},
  year = {2011},
  month = nov
}

@article{Verscharen2013,
  title = {THE DISPERSION RELATIONS AND INSTABILITY THRESHOLDS OF OBLIQUE PLASMA MODES IN THE PRESENCE OF AN ION BEAM},
  volume = {764},
  ISSN = {1538-4357},
  url = {http://dx.doi.org/10.1088/0004-637X/764/1/88},
  DOI = {10.1088/0004-637x/764/1/88},
  number = {1},
  journal = {The Astrophysical Journal},
  publisher = {American Astronomical Society},
  author = {Verscharen,  Daniel and Chandran,  Benjamin D. G.},
  year = {2013},
  month = jan,
  pages = {88}
}

@article{Zhu2023,
  title = {Non-field-aligned Proton Beams and Their Roles in the Growth of Fast Magnetosonic/Whistler Waves: Solar Orbiter Observations},
  volume = {953},
  ISSN = {1538-4357},
  url = {http://dx.doi.org/10.3847/1538-4357/acdc17},
  DOI = {10.3847/1538-4357/acdc17},
  number = {2},
  journal = {The Astrophysical Journal},
  publisher = {American Astronomical Society},
  author = {Zhu,  Xingyu and He,  Jiansen and Duan,  Die and Verscharen,  Daniel and Owen,  Christopher J. and Fedorov,  Andrey and Louarn,  Philippe and Horbury,  Timothy S.},
  year = {2023},
  month = aug,
  pages = {161}
}

@article{Verniero2022,
  title = {Strong Perpendicular Velocity-space Diffusion in Proton Beams Observed by Parker Solar Probe},
  volume = {924},
  ISSN = {1538-4357},
  url = {http://dx.doi.org/10.3847/1538-4357/ac36d5},
  DOI = {10.3847/1538-4357/ac36d5},
  number = {2},
  journal = {The Astrophysical Journal},
  publisher = {American Astronomical Society},
  author = {Verniero,  J. L. and Chandran,  B. D. G. and Larson,  D. E. and Paulson,  K. and Alterman,  B. L. and Badman,  S. and Bale,  S. D. and Bonnell,  J. W. and Bowen,  T. A. and de Wit,  T. Dudok and Kasper,  J. C. and Klein,  K. G. and Lichko,  E. and Livi,  R. and McManus,  M. D. and Rahmati,  A. and Verscharen,  D. and Walters,  J. and Whittlesey,  P. L.},
  year = {2022},
  month = jan,
  pages = {112}
}

@article{Pezzini2024,
  title = {Fully Kinetic Simulations of Proton-beam-driven Instabilities from Parker Solar Probe Observations},
  volume = {975},
  ISSN = {1538-4357},
  url = {http://dx.doi.org/10.3847/1538-4357/ad7465},
  DOI = {10.3847/1538-4357/ad7465},
  number = {1},
  journal = {The Astrophysical Journal},
  publisher = {American Astronomical Society},
  author = {Pezzini,  L. and Zhukov,  A. N. and Bacchini,  F. and Arrò,  G. and López,  R. A. and Micera,  A. and Innocenti,  M. E. and Lapenta,  G.},
  year = {2024},
  month = oct,
  pages = {37}
}

@article{Neugebauer1990,
  title = {A search for evidence of the evolution of rotational discontinuities in the solar wind from nonlinear Alfvén waves},
  volume = {95},
  ISSN = {0148-0227},
  url = {http://dx.doi.org/10.1029/JA095iA01p00013},
  DOI = {10.1029/ja095ia01p00013},
  number = {A1},
  journal = {Journal of Geophysical Research: Space Physics},
  publisher = {American Geophysical Union (AGU)},
  author = {Neugebauer,  M. and Buti,  B.},
  year = {1990},
  month = jan,
  pages = {13–20}
}

@article{Neugebauer2006,
  title = {Comment on the abundances of rotational and tangential discontinuities in the solar wind},
  volume = {111},
  ISSN = {0148-0227},
  url = {http://dx.doi.org/10.1029/2005JA011497},
  DOI = {10.1029/2005ja011497},
  number = {A4},
  journal = {Journal of Geophysical Research: Space Physics},
  publisher = {American Geophysical Union (AGU)},
  author = {Neugebauer,  M.},
  year = {2006},
  month = apr 
}

@article{Livi1987,
  title = {Generation of solar wind proton tails and double beams by Coulomb collisions},
  volume = {92},
  ISSN = {0148-0227},
  url = {http://dx.doi.org/10.1029/JA092iA07p07255},
  DOI = {10.1029/ja092ia07p07255},
  number = {A7},
  journal = {Journal of Geophysical Research: Space Physics},
  publisher = {American Geophysical Union (AGU)},
  author = {Livi,  S. and Marsch,  E.},
  year = {1987},
  month = jul,
  pages = {7255–7261}
}

@article{Tu2002,
  title = {Formation of the proton beam distribution in high‐speed solar wind},
  volume = {107},
  ISSN = {0148-0227},
  url = {http://dx.doi.org/10.1029/2002JA009264},
  DOI = {10.1029/2002ja009264},
  number = {A10},
  journal = {Journal of Geophysical Research: Space Physics},
  publisher = {American Geophysical Union (AGU)},
  author = {Tu,  C.‐Y. and Wang,  L.‐H. and Marsch,  E.},
  year = {2002},
  month = oct 
}

@article{Araneda2008,
  title = {Proton Core Heating and Beam Formation via Parametrically Unstable Alfvén-Cyclotron Waves},
  volume = {100},
  ISSN = {1079-7114},
  url = {http://dx.doi.org/10.1103/PhysRevLett.100.125003},
  DOI = {10.1103/physrevlett.100.125003},
  number = {12},
  journal = {Physical Review Letters},
  publisher = {American Physical Society (APS)},
  author = {Araneda,  Jaime A. and Marsch,  Eckart and F.-Viñas,  Adolfo},
  year = {2008},
  month = mar 
}

@article{Osmane2010,
  title = {On the generation of proton beams in fast solar wind in the presence of obliquely propagating Alfvén waves},
  volume = {115},
  ISSN = {0148-0227},
  url = {http://dx.doi.org/10.1029/2009JA014655},
  DOI = {10.1029/2009ja014655},
  number = {A5},
  journal = {Journal of Geophysical Research: Space Physics},
  publisher = {American Geophysical Union (AGU)},
  author = {Osmane,  A. and Hamza,  A. M. and Meziane,  K.},
  year = {2010},
  month = may 
}

@article{Durovcova2021,
  title = {Proton Beam Abundance Variations and Their Relation to Alpha Particle Properties},
  volume = {923},
  ISSN = {1538-4357},
  url = {http://dx.doi.org/10.3847/1538-4357/ac2c03},
  DOI = {10.3847/1538-4357/ac2c03},
  number = {2},
  journal = {The Astrophysical Journal},
  publisher = {American Astronomical Society},
  author = {Ďurovcová,  Tereza and Šafránková,  Jana and Němeček,  Zdeněk},
  year = {2021},
  month = dec,
  pages = {170}
}

@article{Verniero2020,
  title = {Parker Solar Probe Observations of Proton Beams Simultaneous with Ion-scale Waves},
  volume = {248},
  ISSN = {1538-4365},
  url = {http://dx.doi.org/10.3847/1538-4365/ab86af},
  DOI = {10.3847/1538-4365/ab86af},
  number = {1},
  journal = {The Astrophysical Journal Supplement Series},
  publisher = {American Astronomical Society},
  author = {Verniero,  J. L. and Larson,  D. E. and Livi,  R. and Rahmati,  A. and McManus,  M. D. and Pyakurel,  P. Sharma and Klein,  K. G. and Bowen,  T. A. and Bonnell,  J. W. and Alterman,  B. L. and Whittlesey,  P. L. and Malaspina,  David M. and Bale,  S. D. and Kasper,  J. C. and Case,  A. W. and Goetz,  K. and Harvey,  P. R. and Korreck,  K. E. and MacDowall,  R. J. and Pulupa,  M. and Stevens,  M. L. and de Wit,  T. Dudok},
  year = {2020},
  month = apr,
  pages = {5}
}

@article{Lavraud2021,
  title = {Magnetic reconnection as a mechanism to produce multiple thermal proton populations and beams locally in the solar wind},
  volume = {656},
  ISSN = {1432-0746},
  url = {http://dx.doi.org/10.1051/0004-6361/202141149},
  DOI = {10.1051/0004-6361/202141149},
  journal = {Astronomy \& Astrophysics},
  publisher = {EDP Sciences},
  author = {Lavraud,  B. and Kieokaew,  R. and Fargette,  N. and Louarn,  P. and Fedorov,  A. and André,  N. and Fruit,  G. and Génot,  V. and Réville,  V. and Rouillard,  A. P. and Plotnikov,  I. and Penou,  E. and Barthe,  A. and Prech,  L. and Owen,  C. J. and Bruno,  R. and Allegrini,  F. and Berthomier,  M. and Kataria,  D. and Livi,  S. and Raines,  J. M. and D’Amicis,  R. and Eastwood,  J. P. and Froment,  C. and Laker,  R. and Maksimovic,  M. and Marcucci,  F. and Perri,  S. and Perrone,  D. and Phan,  T. D. and Stansby,  D. and Stawarz,  J. and Toledo-Redondo,  S. and Vaivads,  A. and Verscharen,  D. and Zouganelis,  I. and Angelini,  V. and Evans,  V. and Horbury,  T. S. and O’Brien,  H.},
  year = {2021},
  month = dec,
  pages = {A37}
}

@article{Vasquez2001,
  title = {Evolution and dissipation of imbedded rotational discontinuities and Alfvén waves in nonuniform plasma and the resultant proton heating},
  volume = {106},
  ISSN = {0148-0227},
  url = {http://dx.doi.org/10.1029/2000JA000268},
  DOI = {10.1029/2000ja000268},
  number = {A4},
  journal = {Journal of Geophysical Research: Space Physics},
  publisher = {American Geophysical Union (AGU)},
  author = {Vasquez,  Bernard J. and Hollweg,  Joseph V.},
  year = {2001},
  month = apr,
  pages = {5661–5681}
}

@book{landau1984electrodynamics,
  author    = {Landau, L.D. and Lifshitz, E.M.},
  title     = {Electrodynamics of Continuous Media},
  publisher = {Pergamon Press},
  year      = {1984},
  address   = {England},
  edition   = {2nd}
}

@article{Tsurutani1995,
  title = {Interplanetary discontinuities and Alfv\'en waves},
  volume = {72},
  ISSN = {1572-9672},
  url = {http://dx.doi.org/10.1007/BF00768781},
  DOI = {10.1007/bf00768781},
  number = {1–2},
  journal = {Space Science Reviews},
  publisher = {Springer Science and Business Media LLC},
  author = {Tsurutani,  Bruce T. and Smith,  Edward J. and Ho,  Christian M. and Neugebauer,  Marcia and Goldstein,  Bruce E. and Mok,  John S. and Balogh,  Andre and Southwood,  David and Feldman,  William C.},
  year = {1995},
  month = apr,
  pages = {205–210}
}

@article{Yang2015,
  title = {THE FORMATION OF ROTATIONAL DISCONTINUITIES IN COMPRESSIVE THREE-DIMENSIONAL MHD TURBULENCE},
  volume = {809},
  ISSN = {1538-4357},
  url = {http://dx.doi.org/10.1088/0004-637X/809/2/155},
  DOI = {10.1088/0004-637x/809/2/155},
  number = {2},
  journal = {The Astrophysical Journal},
  publisher = {American Astronomical Society},
  author = {Yang,  Liping and Zhang,  Lei and He,  Jiansen and Tu,  Chuanyi and Wang,  Linghua and Marsch,  Eckart and Wang,  Xin and Zhang,  Shaohua and Feng,  Xueshang},
  year = {2015},
  month = aug,
  pages = {155}
}

@article{Lin1994,
  title = {Structure of reconnection layers in the magnetosphere},
  volume = {65},
  ISSN = {1572-9672},
  url = {http://dx.doi.org/10.1007/BF00749762},
  DOI = {10.1007/bf00749762},
  number = {1–2},
  journal = {Space Science Reviews},
  publisher = {Springer Science and Business Media LLC},
  author = {Lin,  Y. and Lee,  L. C.},
  year = {1994},
  pages = {59–179}
}

@article{Liu2011,
  title = {The effects of strong temperature anisotropy on the kinetic structure of collisionless slow shocks and reconnection exhausts. II. Theory},
  volume = {18},
  ISSN = {1089-7674},
  url = {http://dx.doi.org/10.1063/1.3627147},
  DOI = {10.1063/1.3627147},
  number = {9},
  journal = {Physics of Plasmas},
  publisher = {AIP Publishing},
  author = {Liu,  Yi-Hsin and Drake,  J. F. and Swisdak,  M.},
  year = {2011},
  month = sep 
}

@article{Innocenti2015,
  title = {EVIDENCE OF MAGNETIC FIELD SWITCH-OFF IN COLLISIONLESS MAGNETIC RECONNECTION},
  volume = {810},
  ISSN = {2041-8213},
  url = {http://dx.doi.org/10.1088/2041-8205/810/2/L19},
  DOI = {10.1088/2041-8205/810/2/l19},
  number = {2},
  journal = {The Astrophysical Journal},
  publisher = {American Astronomical Society},
  author = {Innocenti,  M. E. and Goldman,  M. and Newman,  D. and Markidis,  S. and Lapenta,  G.},
  year = {2015},
  month = sep,
  pages = {L19}
}

@article{Lin2009,
  title = {A possible generation mechanism of interplanetary rotational discontinuities},
  volume = {114},
  ISSN = {0148-0227},
  url = {http://dx.doi.org/10.1029/2008JA014008},
  DOI = {10.1029/2008ja014008},
  number = {A8},
  journal = {Journal of Geophysical Research: Space Physics},
  publisher = {American Geophysical Union (AGU)},
  author = {Lin,  C. C. and Tsai,  C. L. and Chen,  H. J. and Weng,  C. J. and Chao,  J. K. and Lee,  L. C.},
  year = {2009},
  month = aug 
}

@article{Le2023Hybrid,
  title = {Hybrid-VPIC: An open-source kinetic/fluid hybrid particle-in-cell code},
  volume = {30},
  ISSN = {1089-7674},
  url = {http://dx.doi.org/10.1063/5.0146529},
  DOI = {10.1063/5.0146529},
  number = {6},
  journal = {Physics of Plasmas},
  publisher = {AIP Publishing},
  author = {Le,  Ari and Stanier,  Adam and Yin,  Lin and Wetherton,  Blake and Keenan,  Brett and Albright,  Brian},
  year = {2023},
  month = jun 
}

@article{Richter1989,
  title = {On the stability of rotational discontinuities},
  volume = {16},
  ISSN = {1944-8007},
  url = {http://dx.doi.org/10.1029/GL016i011p01257},
  DOI = {10.1029/gl016i011p01257},
  number = {11},
  journal = {Geophysical Research Letters},
  publisher = {American Geophysical Union (AGU)},
  author = {Richter, Peter and Scholer,  Manfred},
  year = {1989},
  month = nov,
  pages = {1257–1260}
}

@article{KraussVarban1993,
  title = {Structure and length scales of rotational discontinuities},
  volume = {98},
  ISSN = {0148-0227},
  url = {http://dx.doi.org/10.1029/92JA02362},
  DOI = {10.1029/92ja02362},
  number = {A3},
  journal = {Journal of Geophysical Research: Space Physics},
  publisher = {American Geophysical Union (AGU)},
  author = {Krauss‐Varban,  D.},
  year = {1993},
  month = mar,
  pages = {3907–3917}
}

@article{KraussVarban1995,
  title = {Kinetic structure of rotational discontinuities: Implications for the magnetopause},
  volume = {100},
  ISSN = {0148-0227},
  url = {http://dx.doi.org/10.1029/94JA03034},
  DOI = {10.1029/94ja03034},
  number = {A7},
  journal = {Journal of Geophysical Research: Space Physics},
  publisher = {American Geophysical Union (AGU)},
  author = {Krauss‐Varban,  D. and Karimabadi,  H. and Omidi,  N.},
  year = {1995},
  month = jul,
  pages = {11981–11999}
}

@article{Karimabadi1995,
  title = {On the stability of rotational discontinuities: One‐ and two‐dimensional hybrid simulations},
  volume = {22},
  ISSN = {1944-8007},
  url = {http://dx.doi.org/10.1029/95GL02887},
  DOI = {10.1029/95gl02887},
  number = {21},
  journal = {Geophysical Research Letters},
  publisher = {American Geophysical Union (AGU)},
  author = {Karimabadi,  H. and Krauss‐Varban,  D. and Omidi,  N.},
  year = {1995},
  month = nov,
  pages = {2989–2992}
}

@article{Luo2022,
  title = {Coherence of Ion Cyclotron Resonance in Damped Ion Cyclotron Waves in Space Plasmas},
  volume = {928},
  ISSN = {1538-4357},
  url = {http://dx.doi.org/10.3847/1538-4357/ac52a9},
  DOI = {10.3847/1538-4357/ac52a9},
  number = {1},
  journal = {The Astrophysical Journal},
  publisher = {American Astronomical Society},
  author = {Luo,  Qiaowen and Zhu,  Xingyu and He,  Jiansen and Cui,  Jun and Lai,  Hairong and Verscharen,  Daniel and Duan,  Die},
  year = {2022},
  month = mar,
  pages = {36}
}

@article{Fox2015,
  title = {The Solar Probe Plus Mission: Humanity’s First Visit to Our Star},
  volume = {204},
  ISSN = {1572-9672},
  url = {http://dx.doi.org/10.1007/s11214-015-0211-6},
  DOI = {10.1007/s11214-015-0211-6},
  number = {1–4},
  journal = {Space Science Reviews},
  publisher = {Springer Science and Business Media LLC},
  author = {Fox,  N. J. and Velli,  M. C. and Bale,  S. D. and Decker,  R. and Driesman,  A. and Howard,  R. A. and Kasper,  J. C. and Kinnison,  J. and Kusterer,  M. and Lario,  D. and Lockwood,  M. K. and McComas,  D. J. and Raouafi,  N. E. and Szabo,  A.},
  year = {2015},
  month = nov,
  pages = {7–48}
}

@article{Mueller2020,
  title = {The Solar Orbiter mission: Science overview},
  volume = {642},
  ISSN = {1432-0746},
  url = {http://dx.doi.org/10.1051/0004-6361/202038467},
  DOI = {10.1051/0004-6361/202038467},
  journal = {Astronomy \& Astrophysics},
  publisher = {EDP Sciences},
  author = {M\"{u}ller,  D. and St. Cyr,  O. C. and Zouganelis,  I. and Gilbert,  H. R. and Marsden,  R. and Nieves-Chinchilla,  T. and Antonucci,  E. and Auchère,  F. and Berghmans,  D. and Horbury,  T. S. and Howard,  R. A. and Krucker,  S. and Maksimovic,  M. and Owen,  C. J. and Rochus,  P. and Rodriguez-Pacheco,  J. and Romoli,  M. and Solanki,  S. K. and Bruno,  R. and Carlsson,  M. and Fludra,  A. and Harra,  L. and Hassler,  D. M. and Livi,  S. and Louarn,  P. and Peter,  H. and Sch\"{u}hle,  U. and Teriaca,  L. and del Toro Iniesta,  J. C. and Wimmer-Schweingruber,  R. F. and Marsch,  E. and Velli,  M. and De Groof,  A. and Walsh,  A. and Williams,  D.},
  year = {2020},
  month = sep,
  pages = {A1}
}

@article{Kasper2015,
  title = {Solar Wind Electrons Alphas and Protons (SWEAP) Investigation: Design of the Solar Wind and Coronal Plasma Instrument Suite for Solar Probe Plus},
  volume = {204},
  ISSN = {1572-9672},
  url = {http://dx.doi.org/10.1007/s11214-015-0206-3},
  DOI = {10.1007/s11214-015-0206-3},
  number = {1–4},
  journal = {Space Science Reviews},
  publisher = {Springer Science and Business Media LLC},
  author = {Kasper,  Justin C. and Abiad,  Robert and Austin,  Gerry and Balat-Pichelin,  Marianne and Bale,  Stuart D. and Belcher,  John W. and Berg,  Peter and Bergner,  Henry and Berthomier,  Matthieu and Bookbinder,  Jay and Brodu,  Etienne and Caldwell,  David and Case,  Anthony W. and Chandran,  Benjamin D. G. and Cheimets,  Peter and Cirtain,  Jonathan W. and Cranmer,  Steven R. and Curtis,  David W. and Daigneau,  Peter and Dalton,  Greg and Dasgupta,  Brahmananda and DeTomaso,  David and Diaz-Aguado,  Millan and Djordjevic,  Blagoje and Donaskowski,  Bill and Effinger,  Michael and Florinski,  Vladimir and Fox,  Nichola and Freeman,  Mark and Gallagher,  Dennis and Gary,  S. Peter and Gauron,  Tom and Gates,  Richard and Goldstein,  Melvin and Golub,  Leon and Gordon,  Dorothy A. and Gurnee,  Reid and Guth,  Giora and Halekas,  Jasper and Hatch,  Ken and Heerikuisen,  Jacob and Ho,  George and Hu,  Qiang and Johnson,  Greg and Jordan,  Steven P. and Korreck,  Kelly E. and Larson,  Davin and Lazarus,  Alan J. and Li,  Gang and Livi,  Roberto and Ludlam,  Michael and Maksimovic,  Milan and McFadden,  James P. and Marchant,  William and Maruca,  Bennet A. and McComas,  David J. and Messina,  Luciana and Mercer,  Tony and Park,  Sang and Peddie,  Andrew M. and Pogorelov,  Nikolai and Reinhart,  Matthew J. and Richardson,  John D. and Robinson,  Miles and Rosen,  Irene and Skoug,  Ruth M. and Slagle,  Amanda and Steinberg,  John T. and Stevens,  Michael L. and Szabo,  Adam and Taylor,  Ellen R. and Tiu,  Chris and Turin,  Paul and Velli,  Marco and Webb,  Gary and Whittlesey,  Phyllis and Wright,  Ken and Wu,  S. T. and Zank,  Gary},
  year = {2015},
  month = oct,
  pages = {131–186}
}

@article{Livi2022,
  title = {The Solar Probe ANalyzer—Ions on the Parker Solar Probe},
  volume = {938},
  ISSN = {1538-4357},
  url = {http://dx.doi.org/10.3847/1538-4357/ac93f5},
  DOI = {10.3847/1538-4357/ac93f5},
  number = {2},
  journal = {The Astrophysical Journal},
  publisher = {American Astronomical Society},
  author = {Livi,  Roberto and Larson,  Davin E. and Kasper,  Justin C. and Abiad,  Robert and Case,  A. W. and Klein,  Kristopher G. and Curtis,  David W. and Dalton,  Gregory and Stevens,  Michael and Korreck,  Kelly E. and Ho,  George and Robinson,  Miles and Tiu,  Chris and Whittlesey,  Phyllis L. and Verniero,  Jaye L. and Halekas,  Jasper and McFadden,  James and Marckwordt,  Mario and Slagle,  Amanda and Abatcha,  Mamuda and Rahmati,  Ali and McManus,  Michael D.},
  year = {2022},
  month = oct,
  pages = {138}
}

@article{Owen2020,
  title = {The Solar Orbiter Solar Wind Analyser (SWA) suite},
  volume = {642},
  ISSN = {1432-0746},
  url = {http://dx.doi.org/10.1051/0004-6361/201937259},
  DOI = {10.1051/0004-6361/201937259},
  journal = {Astronomy \& Astrophysics},
  publisher = {EDP Sciences},
  author = {Owen,  C. J. and Bruno,  R. and Livi,  S. and Louarn,  P. and Al Janabi,  K. and Allegrini,  F. and Amoros,  C. and Baruah,  R. and Barthe,  A. and Berthomier,  M. and Bordon,  S. and Brockley-Blatt,  C. and Brysbaert,  C. and Capuano,  G. and Collier,  M. and DeMarco,  R. and Fedorov,  A. and Ford,  J. and Fortunato,  V. and Fratter,  I. and Galvin,  A. B. and Hancock,  B. and Heirtzler,  D. and Kataria,  D. and Kistler,  L. and Lepri,  S. T. and Lewis,  G. and Loeffler,  C. and Marty,  W. and Mathon,  R. and Mayall,  A. and Mele,  G. and Ogasawara,  K. and Orlandi,  M. and Pacros,  A. and Penou,  E. and Persyn,  S. and Petiot,  M. and Phillips,  M. and Přech,  L. and Raines,  J. M. and Reden,  M. and Rouillard,  A. P. and Rousseau,  A. and Rubiella,  J. and Seran,  H. and Spencer,  A. and Thomas,  J. W. and Trevino,  J. and Verscharen,  D. and Wurz,  P. and Alapide,  A. and Amoruso,  L. and André,  N. and Anekallu,  C. and Arciuli,  V. and Arnett,  K. L. and Ascolese,  R. and Bancroft,  C. and Bland,  P. and Brysch,  M. and Calvanese,  R. and Castronuovo,  M. and Čermák,  I. and Chornay,  D. and Clemens,  S. and Coker,  J. and Collinson,  G. and D’Amicis,  R. and Dandouras,  I. and Darnley,  R. and Davies,  D. and Davison,  G. and De Los Santos,  A. and Devoto,  P. and Dirks,  G. and Edlund,  E. and Fazakerley,  A. and Ferris,  M. and Frost,  C. and Fruit,  G. and Garat,  C. and Génot,  V. and Gibson,  W. and Gilbert,  J. A. and de Giosa,  V. and Gradone,  S. and Hailey,  M. and Horbury,  T. S. and Hunt,  T. and Jacquey,  C. and Johnson,  M. and Lavraud,  B. and Lawrenson,  A. and Leblanc,  F. and Lockhart,  W. and Maksimovic,  M. and Malpus,  A. and Marcucci,  F. and Mazelle,  C. and Monti,  F. and Myers,  S. and Nguyen,  T. and Rodriguez-Pacheco,  J. and Phillips,  I. and Popecki,  M. and Rees,  K. and Rogacki,  S. A. and Ruane,  K. and Rust,  D. and Salatti,  M. and Sauvaud,  J. A. and Stakhiv,  M. O. and Stange,  J. and Stubbs,  T. and Taylor,  T. and Techer,  J.-D. and Terrier,  G. and Thibodeaux,  R. and Urdiales,  C. and Varsani,  A. and Walsh,  A. P. and Watson,  G. and Wheeler,  P. and Willis,  G. and Wimmer-Schweingruber,  R. F. and Winter,  B. and Yardley,  J. and Zouganelis,  I.},
  year = {2020},
  month = sep,
  pages = {A16}
}

@article{Klein2021,
  title = {Inferred Linear Stability of Parker Solar Probe Observations Using One- and Two-component Proton Distributions},
  volume = {909},
  ISSN = {1538-4357},
  url = {http://dx.doi.org/10.3847/1538-4357/abd7a0},
  DOI = {10.3847/1538-4357/abd7a0},
  number = {1},
  journal = {The Astrophysical Journal},
  publisher = {American Astronomical Society},
  author = {Klein,  K. G. and Verniero,  J. L. and Alterman,  B. and Bale,  S. and Case,  A. and Kasper,  J. C. and Korreck,  K. and Larson,  D. and Lichko,  E. and Livi,  R. and McManus,  M. and Martinović,  M. and Rahmati,  A. and Stevens,  M. and Whittlesey,  P.},
  year = {2021},
  month = mar,
  pages = {7}
}

@article{Louarn2021,
  title = {Multiscale views of an Alfvénic slow solar wind: 3D velocity distribution functions observed by the Proton-Alpha Sensor of Solar Orbiter},
  volume = {656},
  ISSN = {1432-0746},
  url = {http://dx.doi.org/10.1051/0004-6361/202141095},
  DOI = {10.1051/0004-6361/202141095},
  journal = {Astronomy \& Astrophysics},
  publisher = {EDP Sciences},
  author = {Louarn,  P. and Fedorov,  A. and Prech,  L. and Owen,  C. J. and Bruno,  R. and Livi,  S. and Lavraud,  B. and Rouillard,  A. P. and Génot,  V. and André,  N. and Fruit,  G. and Réville,  V. and Kieokaew,  R. and Plotnikov,  I. and Penou,  E. and Barthe,  A. and Khataria,  D. and Berthomier,  M. and D’Amicis,  R. and Sorriso-Valvo,  L. and Allegrini,  F. and Raines,  J. and Verscharen,  D. and Fortunato,  V. and Mele,  G. and Horbury,  T. S. and O’brien,  H. and Evans,  V. and Angelini,  V. and Maksimovic,  M. and Kasper,  J. C. and Bale,  S. D.},
  year = {2021},
  month = dec,
  pages = {A36}
}

@article{Shen2024,
  title = {Comparing Plasma Anisotropy Associated with Solar Wind Discontinuities and Alfvénic Fluctuations},
  volume = {961},
  ISSN = {1538-4357},
  url = {http://dx.doi.org/10.3847/1538-4357/ad110b},
  DOI = {10.3847/1538-4357/ad110b},
  number = {1},
  journal = {The Astrophysical Journal},
  publisher = {American Astronomical Society},
  author = {Shen,  Yangyang and Artemyev,  Anton and Angelopoulos,  Vassilis and Liu,  Terry Z. and Vasko,  Ivan},
  year = {2024},
  month = jan,
  pages = {41}
}

@article{Bale2019,
  title = {Highly structured slow solar wind emerging from an equatorial coronal hole},
  volume = {576},
  ISSN = {1476-4687},
  url = {http://dx.doi.org/10.1038/s41586-019-1818-7},
  DOI = {10.1038/s41586-019-1818-7},
  number = {7786},
  journal = {Nature},
  publisher = {Springer Science and Business Media LLC},
  author = {Bale,  S. D. and Badman,  S. T. and Bonnell,  J. W. and Bowen,  T. A. and Burgess,  D. and Case,  A. W. and Cattell,  C. A. and Chandran,  B. D. G. and Chaston,  C. C. and Chen,  C. H. K. and Drake,  J. F. and de Wit,  T. Dudok and Eastwood,  J. P. and Ergun,  R. E. and Farrell,  W. M. and Fong,  C. and Goetz,  K. and Goldstein,  M. and Goodrich,  K. A. and Harvey,  P. R. and Horbury,  T. S. and Howes,  G. G. and Kasper,  J. C. and Kellogg,  P. J. and Klimchuk,  J. A. and Korreck,  K. E. and Krasnoselskikh,  V. V. and Krucker,  S. and Laker,  R. and Larson,  D. E. and MacDowall,  R. J. and Maksimovic,  M. and Malaspina,  D. M. and Martinez-Oliveros,  J. and McComas,  D. J. and Meyer-Vernet,  N. and Moncuquet,  M. and Mozer,  F. S. and Phan,  T. D. and Pulupa,  M. and Raouafi,  N. E. and Salem,  C. and Stansby,  D. and Stevens,  M. and Szabo,  A. and Velli,  M. and Woolley,  T. and Wygant,  J. R.},
  year = {2019},
  month = dec,
  pages = {237–242}
}

@article{Kasper2019,
  title = {Alfvénic velocity spikes and rotational flows in the near-Sun solar wind},
  volume = {576},
  ISSN = {1476-4687},
  url = {http://dx.doi.org/10.1038/s41586-019-1813-z},
  DOI = {10.1038/s41586-019-1813-z},
  number = {7786},
  journal = {Nature},
  publisher = {Springer Science and Business Media LLC},
  author = {Kasper,  J. C. and Bale,  S. D. and Belcher,  J. W. and Berthomier,  M. and Case,  A. W. and Chandran,  B. D. G. and Curtis,  D. W. and Gallagher,  D. and Gary,  S. P. and Golub,  L. and Halekas,  J. S. and Ho,  G. C. and Horbury,  T. S. and Hu,  Q. and Huang,  J. and Klein,  K. G. and Korreck,  K. E. and Larson,  D. E. and Livi,  R. and Maruca,  B. and Lavraud,  B. and Louarn,  P. and Maksimovic,  M. and Martinovic,  M. and McGinnis,  D. and Pogorelov,  N. V. and Richardson,  J. D. and Skoug,  R. M. and Steinberg,  J. T. and Stevens,  M. L. and Szabo,  A. and Velli,  M. and Whittlesey,  P. L. and Wright,  K. H. and Zank,  G. P. and MacDowall,  R. J. and McComas,  D. J. and McNutt,  R. L. and Pulupa,  M. and Raouafi,  N. E. and Schwadron,  N. A.},
  year = {2019},
  month = dec,
  pages = {228–231}
}

@article{Larosa2021,
  title = {Switchbacks: statistical properties and deviations from Alfvénicity},
  volume = {650},
  ISSN = {1432-0746},
  url = {http://dx.doi.org/10.1051/0004-6361/202039442},
  DOI = {10.1051/0004-6361/202039442},
  journal = {Astronomy \& Astrophysics},
  publisher = {EDP Sciences},
  author = {Larosa,  A. and Krasnoselskikh,  V. and Dudok de Wit,  T. and Agapitov,  O. and Froment,  C. and Jagarlamudi,  V. K. and Velli,  M. and Bale,  S. D. and Case,  A. W. and Goetz,  K. and Harvey,  P. and Kasper,  J. C. and Korreck,  K. E. and Larson,  D. E. and MacDowall,  R. J. and Malaspina,  D. and Pulupa,  M. and Revillet,  C. and Stevens,  M. L.},
  year = {2021},
  month = jun,
  pages = {A3}
}

@article{AkhavanTafti2022,
  title = {Magnetic Switchbacks Heat the Solar Corona},
  volume = {937},
  ISSN = {2041-8213},
  url = {http://dx.doi.org/10.3847/2041-8213/ac913d},
  DOI = {10.3847/2041-8213/ac913d},
  number = {2},
  journal = {The Astrophysical Journal Letters},
  publisher = {American Astronomical Society},
  author = {Akhavan-Tafti,  M. and Kasper,  J. and Huang,  J. and Thomas,  L.},
  year = {2022},
  month = oct,
  pages = {L39}
}

@article{Bizien2023,
  title = {Are Switchback Boundaries Observed by Parker Solar Probe Closed?},
  volume = {958},
  ISSN = {1538-4357},
  url = {http://dx.doi.org/10.3847/1538-4357/acf99a},
  DOI = {10.3847/1538-4357/acf99a},
  number = {1},
  journal = {The Astrophysical Journal},
  publisher = {American Astronomical Society},
  author = {Bizien,  Nina and Dudok de Wit,  Thierry and Froment,  Clara and Velli,  Marco and Case,  Anthony W. and Bale,  Stuart D. and Kasper,  Justin and Whittlesey,  Phyllis and MacDowall,  Robert and Larson,  Davin},
  year = {2023},
  month = nov,
  pages = {23}
}

@article{Malara2021,
  title = {Charged-particle chaotic dynamics in rotational discontinuities},
  volume = {104},
  ISSN = {2470-0053},
  url = {http://dx.doi.org/10.1103/PhysRevE.104.025208},
  DOI = {10.1103/physreve.104.025208},
  number = {2},
  journal = {Physical Review E},
  publisher = {American Physical Society (APS)},
  author = {Malara,  Francesco and Perri,  Silvia and Zimbardo,  Gaetano},
  year = {2021},
  month = aug 
}

@article{Egedal2005,
  title = {In SituDiscovery of an Electrostatic Potential,  Trapping Electrons and Mediating Fast Reconnection in the Earth’s Magnetotail},
  volume = {94},
  ISSN = {1079-7114},
  url = {http://dx.doi.org/10.1103/PhysRevLett.94.025006},
  DOI = {10.1103/physrevlett.94.025006},
  number = {2},
  journal = {Physical Review Letters},
  publisher = {American Physical Society (APS)},
  author = {Egedal,  J. and Øieroset,  M. and Fox,  W. and Lin,  R. P.},
  year = {2005},
  month = jan 
}

@article{Hutchinson2020,
  title = {Particle Trapping in Axisymmetric Electron Holes},
  volume = {125},
  ISSN = {2169-9402},
  url = {http://dx.doi.org/10.1029/2020JA028093},
  DOI = {10.1029/2020ja028093},
  number = {8},
  journal = {Journal of Geophysical Research: Space Physics},
  publisher = {American Geophysical Union (AGU)},
  author = {Hutchinson,  I. H.},
  year = {2020},
  month = aug 
}

@article{Xie2024,
  title = {Electron scale coherent structure as micro accelerator in the Earth’s magnetosheath},
  volume = {15},
  ISSN = {2041-1723},
  url = {http://dx.doi.org/10.1038/s41467-024-45040-5},
  DOI = {10.1038/s41467-024-45040-5},
  number = {1},
  journal = {Nature Communications},
  publisher = {Springer Science and Business Media LLC},
  author = {Xie,  Zi-Kang and Zong,  Qiu-Gang and Yue,  Chao and Zhou,  Xu-Zhi and Liu,  Zhi-Yang and He,  Jian-Sen and Hao,  Yi-Xin and Ng,  Chung-Sang and Zhang,  Hui and Yao,  Shu-Tao and Pollock,  Craig and Le,  Guan and Ergun,  Robert and Lindqvist,  Per-Arne},
  year = {2024},
  month = jan 
}

@article{Laker2023,
  title = {Coherent deflection pattern and associated temperature enhancements in the near-Sun solar wind},
  volume = {527},
  ISSN = {1365-2966},
  url = {http://dx.doi.org/10.1093/mnras/stad3351},
  DOI = {10.1093/mnras/stad3351},
  number = {4},
  journal = {Monthly Notices of the Royal Astronomical Society},
  publisher = {Oxford University Press (OUP)},
  author = {Laker,  Ronan and Horbury,  T S and Woodham,  L D and Bale,  S D and Matteini,  L},
  year = {2023},
  month = nov,
  pages = {10440–10447}
}

@techreport{Post1959,
  author       = {Post, R F},
  title        = {SUMMARY OF UCRL PYROTRON (MIRROR MACHINE) PROGRAM},
  institution  = {California.  Univ., Livermore.  Radiation Lab. Page(s):  39},
  url          = {https://www.osti.gov/biblio/4275223},
  place        = {Country unknown/Code not available},
  year         = {1959},
  month        = {10}}

@article{Arro2024,
title = {Large-scale Linear Magnetic Holes with Magnetic Mirror Properties in Hybrid Simulations of Solar Wind Turbulence},
volume = {970},
ISSN = {2041-8213},
url = {http://dx.doi.org/10.3847/2041-8213/ad61da},
DOI = {10.3847/2041-8213/ad61da},
number = {1},
journal = {The Astrophysical Journal Letters},
publisher = {American Astronomical Society},
author = {Arrò, Giuseppe and Califano, Francesco and Pucci, Francesco and Karlsson, Tomas and Li, Hui},
year = {2024},
month = jul,
pages = {L6}
}

@article{Balikhin2009,
title = {THEMIS observations of mirror structures: Magnetic holes and instability threshold},
volume = {36},
ISSN = {1944-8007},
url = {http://dx.doi.org/10.1029/2008GL036923},
DOI = {10.1029/2008gl036923},
number = {3},
journal = {Geophysical Research Letters},
publisher = {American Geophysical Union (AGU)},
author = {Balikhin, M. A. and Sagdeev, R. Z. and Walker, S. N. and Pokhotelov, O. A. and Sibeck, D. G. and Beloff, N. and Dudnikova, G.},
year = {2009},
month = feb
}

@article{Burlaga1977,
  title = {Interplanetary current sheets at 1 AU},
  volume = {82},
  ISSN = {0148-0227},
  url = {http://dx.doi.org/10.1029/JA082i022p03191},
  DOI = {10.1029/ja082i022p03191},
  number = {22},
  journal = {Journal of Geophysical Research},
  publisher = {American Geophysical Union (AGU)},
  author = {Burlaga,  L. F. and Lemaire,  J. F. and Turner,  J. M.},
  year = {1977},
  month = aug,
  pages = {3191–3200}
}

@article{Neugebauer1984,
  title = {A reexamination of rotational and tangential discontinuities in the solar wind},
  volume = {89},
  ISSN = {0148-0227},
  url = {http://dx.doi.org/10.1029/JA089iA07p05395},
  DOI = {10.1029/ja089ia07p05395},
  number = {A7},
  journal = {Journal of Geophysical Research: Space Physics},
  publisher = {American Geophysical Union (AGU)},
  author = {Neugebauer,  M. and Clay,  D. R. and Goldstein,  B. E. and Tsurutani,  B. T. and Zwickl,  R. D.},
  year = {1984},
  month = jul,
  pages = {5395–5408}
}

@inproceedings{boris1970relativistic,
  title={Relativistic plasma simulation-optimization of a hybrid code},
  author={Boris, Jay P and others},
  booktitle={Proc. Fourth Conf. Num. Sim. Plasmas},
  pages={3--67},
  year={1970}
}
\bibliographystyle{aasjournalv7}



\end{document}